\documentclass[journal]{IEEEtran}

\normalsize
\usepackage[dvipsnames]{xcolor}
\usepackage{cite}
\usepackage[dvips]{graphicx}
\ifCLASSOPTIONcompsoc
\usepackage[caption=false,font=normalsize,labelfon
t=sf,textfont=sf]{subfig}
\else
\usepackage[caption=false,font=footnotesize]{subfi
g}
\fi
\usepackage[cmex10]{amsmath}
\usepackage{amsthm,amssymb}
\usepackage{bigints}
\usepackage{multirow}
\usepackage{xcolor}

\newtheorem{theorem}{Theorem}
\newtheorem{lemma}{Lemma}
\newtheorem{remark}{Remark}
\newtheorem{corollary}{Corollary}

\DeclareMathOperator*{\argmax}{arg\,max}

\newcommand{\yet}[1]{\textcolor{red}{#1}}

\begin{document}

\title{Secure Millimeter-Wave Ad Hoc Communications Using Physical Layer Security}
\author{Yuanyu Zhang, \IEEEmembership{Member, IEEE},
		Yulong Shen,  \IEEEmembership{Member, IEEE},
		Xiaohong Jiang, \IEEEmembership{Senior Member, IEEE}
		and Shoji Kasahara, \IEEEmembership{Member, IEEE}
\thanks{Y. Zhang and S. Kasahara are with the Graduate School of Science and Technology, Nara Institute of Science and Technology, 8916-5 Takayama, Ikoma, Nara, 630-0192, Japan. Email: \{yyzhang, kasahara\}@is.naist.jp.}
\thanks{Y. Shen is with the School of Computer Science and Technology, Xidian University, Xi'an, Shaanxi, China. Email: ylshen@mail.xidian.edu.cn.}
\thanks{X. Jiang is with the School of Systems Information Science, Future University Hakodate, Hakodate, Hokkaido, Japan, and also with  the School of Computer Science and Technology, Xidian University, Xi'an, Shaanxi, China. Email: jiang@fun.ac.jp.}
}

\maketitle

\begin{abstract}
Millimeter-wave (mmWave) communications are highly promising to improve the capacity of modern wireless networks, while the physical layer security (PLS) techniques hold great potential to enhance the critical secrecy performance therein. By carefully exploiting the significant signal difference between the Non-Light-of-Sight (NLoS) and Line-of-Sight (LoS) mmWave links, this paper proposes a Sight-based Cooperative Jamming (SCJ) scheme to improve the PLS performance of mmWave ad hoc communications. In this scheme, each potential jammer that has no LoS link to its nearest receiver but may have LoS links to eavesdroppers is selected with a certain probability to generate artificial noise such that channel advantages at legitimate receivers can be achieved. For performance modeling of the new jamming scheme, novel and efficient theoretical approximation approaches are firstly developed to enable the challenging issue of interference distribution modeling to be tackled, and then a theoretical framework based on stochastic geometry is proposed to capture the secrecy transmission capacity behavior under the SCJ scheme. Finally, extensive numerical results are provided to illustrate the SCJ scheme under various network scenarios.
\end{abstract}

\begin{IEEEkeywords}
Physical layer security, millimeter-wave ad hoc communications, cooperative jamming.
\end{IEEEkeywords}

\IEEEpeerreviewmaketitle

\section {Introduction}\label{sec:introduction} 
\IEEEPARstart{T}{he} explosive growth of mobile devices in the past decade poses a significant challenge to the capacity of modern wireless networks.
To tackle this challenge, the wireless industry has advocated shifting wireless communications from the narrow microwave band below $6$ GHz to the extremely wide millimeter-wave (mmWave) band between $30$ GHz and $300$ GHz \cite{Rappaport2013IEEEAccess,Wang2018CST}. 
However, like conventional microwave communications, the secrecy issue of mmWave communications caused by eavesdropping attacks has also been regarded as a major concern \cite{Zou2016IEEEProceedings,ma2018security}. 
Recent research shows that the physical layer security (PLS) technology, which exploits the random physical layer features of wireless channels to achieve information-theoretic security, is highly promising to improve the secrecy of wireless communications \cite{Mukherjee2014,NYang2015,Jameel2018CST,YWu2018JSAC}. 
This paper therefore investigates the PLS issue in mmWave ad hoc communications. In particular, we aim to propose effective PLS schemes for mmWave ad hoc communications and evaluate the PLS performance therein. 

Although the PLS performance of microwave  communications has been extensively studied (e.g., \cite{ChMa2015,YLiu2016TOC,RZhang2016TWC, JWang2018JSAC, Zhang2019TCOM}), the results cannot be directly applied to the mmWave communications due to the intrinsic physical layer features of mmWave channels \cite{Rappaport2017TAP,Sun2018TvT, Shafi2018CoMMag}. 
This motivates researchers to investigate the PLS performance of mmWave communications. 
The authors in \cite{YZhu2017TWC} considered a planar mmWave ad hoc network with multiple transmission pairs and eavesdroppers, and evaluated the average achievable secrecy rate of a single pair. 
In particular, the authors investigated the PLS potentials of an artificial noise (AN)-aided transmission scheme, where each transmitter allocates a fraction of its transmit power for AN transmission, and showed that AN may be ineffective for improving the secrecy rate. 
Apart from planar ad hoc networks, research efforts have also been devoted to the PLS performance evaluation of other mmWave scenarios, such as  3D ad hoc networks with unmanned aerial vehicles (UAVs) \cite{YZhu2018JSAC}, downlink transmissions of cellular networks \cite{WangC2016TWC}, downlink transmissions of hybrid cellular networks \cite{SVuppala2018TCOM, WWang2018TCOM}, where microwave BSs coexist with mmWave BSs, simultaneous wireless information and power transfer (SWIPT) networks \cite{XSun2019TIFS} and simple two-hop relay systems with multiple eavesdroppers \cite{YJu2018TWCSafeguarding, XSun2019WCL}. 
Please refer to the Related Work section (Section \ref{sec:related-work}) for the detailed introduction of these works.

The above works adopted well-known point processes, like the Poisson Point Processes (PPP) \cite{YZhu2017TWC,WangC2016TWC,SVuppala2018TCOM, WWang2018TCOM, XSun2019TIFS} and Matern Hardcore Point Process (MHPP) \cite{YZhu2018JSAC}, to model the locations of nodes, thanks to their ability to enable network-scale performance analysis. 
However, to retain the mathematical tractability of the point processes, these works applied either no PLS schemes or less effective PLS schemes, like the AN-aided transmission in \cite{YZhu2017TWC,WangC2016TWC} and cooperative jamming in \cite{YZhu2018JSAC}, where part of the transmitters send artificial noise that is \emph{equally detrimental} to legitimate receivers and eavesdroppers. 
These PLS schemes are based on the design disciplines for microwave communications, while we believe that effective PLS schemes specific to mmWave communications should carefully exploit the distinctive physical layer features of mmWave channels. 
For instance, Non-Line-of-Sight (NLoS) mmWave links suffer from much more severe signal attenuation than LoS mmWave links \cite{Rappaport2017TAP,Sun2018TvT, Shafi2018CoMMag}. 
Thus, cooperative jamming schemes can use nodes that have no LoS links (i.e., have NLoS links) to legitimate receivers but have LoS links to eavesdroppers to cause less interference at the former than at the latter, thus improving the PLS performance.
Unfortunately, to the best of our knowledge, such appealing idea of designing PLS schemes based on \yet{intrinsic physical layer features of mmWave channels (e.g., sensitive to blockage, significantly different attenuation between LoS links and NLoS links)} was ignored in previous works. 
This motivates our PLS scheme design and the corresponding PLS performance evaluation in this paper.

This paper considers an mmWave ad hoc network where the locations of transmission pairs, \emph{potential} jammers and eavesdroppers are modeled by PPPs. 
The main contributions of this paper can be summarized as follows.
\begin{itemize} 
\item By carefully exploiting the significant signal difference between NLoS and LoS mmWave links, we propose a novel Sight-based Cooperative Jamming (SCJ) scheme to improve the PLS performance of the transmission pairs. 
With the aim to achieve channel advantages for the legitimate receivers, in the SCJ scheme, each potential jammer that has no LoS link to its \emph{nearest} receiver but may have LoS links to the eavesdroppers is selected as a jammer with a certain probability to generate artificial noise. 
\yet{To the best of our knowledge, this paper represents the first work that exploits the intrinsic physical layer features of mmWave channels in the design of PLS schemes for mmWave networks.  It is expected that this work will shed light on a new approach for the design of secure mmWave communication systems. }
\item We adopt the secrecy transmission capacity (STC) \cite{Zhou2011}, i.e., the average sum rate of transmissions in perfect secrecy per unit area, as the metric to investigate the PLS performance of the network under the new jamming scheme. 
\yet{Existing works mainly assume that jammers follow homogeneous PPPs, while the jammers in this paper follow Poisson Hole Processes (PHPs) and the resultant regions where jammers reside have irregular shapes, posing a significant challenge to the interference modeling for legitimate receivers and eavesdroppers.
 To tackle this challenge, we apply region approximation to }
develop novel and efficient theoretical approaches to approximate the inhomogeneous spatial distribution of the jammers such that the challenging issue of interference distribution modeling can be tackled. 
With the help of the approximations, we then develop a theoretical framework based on stochastic geometry to derive the connection probability and secrecy probability of transmission pairs, based on which the STC analysis is then conducted. 
\yet{The results in this paper show that our new approach for interference modeling under the PHP enables a more accurate analysis to be conducted for both connection probability and secrecy probability.}
\item Finally, we provide extensive numerical results to illustrate the effectiveness of the approximations as well as the optimal STC performances of the proposed SCJ scheme under various network scenarios. 
\yet{The results showed that our scheme can improve the network STC performance and outperforms the typical partial jamming scheme in terms of both STC and energy efficiency performances.}
\end{itemize}

The remainder of this paper is organized as follows. 
Section \ref{sec:related-work} introduces the related work and Section \ref{sec:sys-model} presents the system model as well as the SCJ scheme. 
To facilitate the theoretical analysis, we consider a simplified scenario with only one transmission pair in Section \ref{sec:per-ana-simple-case}. 
The theoretical analysis for the general scenario is presented in Section \ref{sec:per-ana-general-case}. 
We provide numerical results in Section \ref{sec:num-res} and finally conclude the paper in Section \ref{sec:con}.

\section{Related Work}\label{sec:related-work}
Extensive research efforts have been devoted to the study of PLS issue in mmWave communications, which can be roughly classified into two categories depending on whether the network model and performance analysis are based on stochastic geometry or not. 

In stochastic geometry-based works, the locations of legitimate nodes (e.g., transmitters, BSs, receivers) and eavesdroppers are usually modeled by independent and homogenous PPPs to characterize large-scale network scenarios, like ad hoc networks and cellular networks. 
The subsequent description of the related works is based on this prerequisite, unless stated otherwise. 
As introduced in the previous section, the authors in \cite{YZhu2017TWC} considered a planar ad hoc network with multiple transmission pairs and eavesdroppers, and evaluated the average achievable secrecy rate under a simple AN-based transmission scheme, where each transmitter allocates a fraction of its transmit power to radiate AN.  
In \cite{YZhu2018JSAC}, a 3D ad hoc network was considered, where BS-like UAVs, whose locations are modeled by an MHPP rather than a PPP, transmit to ground receivers in the presence of ground eavesdroppers. 
The authors then investigated the secrecy rate performance under a simple cooperative jamming scheme, where part of the UAVs generate AN \emph{irrespective of their link conditions to the ground receivers}. 
Considering the downlink transmission of a cellular network, the authors in \cite{WangC2016TWC} analyzed the secure connectivity probability and average number of perfect links per unit area in the noise-limited scenario under both non-colluding and colluding eavesdropping cases. 
In addition, they also investigated the  average number of perfect links per unit area in the interference-limited scenario under only the non-colluding case and a simple AN scheme similar to the one in \cite{YZhu2017TWC}. 
A hybrid cellular network, where mmWave BSs coexist with microwave BSs, was considered in \cite{SVuppala2018TCOM} and the PLS performances of connection outage probability, secrecy outage probability and secrecy rate were studied. 
The hybrid cellular network scenario was also considered in \cite{WWang2018TCOM}, while, different from the passive eavesdroppers in \cite{SVuppala2018TCOM}, the eavesdroppers here send pilot signals during the channel training phase to improve the quality of intercepted signals.  
Connection and secrecy performances were then investigated under this new attacking scenario. 
The downlink transmissions of an SWIPT cellular network  were examined in \cite{ XSun2019TIFS}, where BSs can simultaneously send information and transfer power to user devices.  
The performances of energy-information coverage probability, secrecy probability and secrecy throughput were also investigated under non-colluding or colluding eavesdropping cases. 

Note that the above works adopted either no PLS schemes or ineffective AN and cooperative jamming schemes, where the noise or jamming signals are equally detrimental to both the legitimate receivers and eavesdroppers, failing  to create channel advantages for the former. 
The main reason is to ensure the mathematical tractability of the point processes (e.g., PPPs and MHPPs), since the adopted schemes lead to homogenous point processes of interfering nodes. 
Differing from these works, this paper proposes an effective SCJ scheme, which carefully exploits the blockage effects between jammers and legitimate receivers to create better legitimate channels.  
As a result, the point process of the jammers becomes inhomogeneous, and thus new and dedicated theoretical analysis is required to model the interference from the jammers in our scenario. 

Apart from the above works, there also exist some other stochastic geometry-based works focusing on multiple-input multiple-output transmissions \cite{YJu2018TWCSafeguarding, WYang2019IEEEAccess}, SWIPT transmissions \cite{XSun2019WCL}, relaying transmissions \cite{RMa2019IEEEAccess} and cognitive radio networks \cite{YSong2019IEEEAccess}. 
However,  different from the above ones, these works considered only one or two source-destination pairs. 
In addition to the stochastic geometry-based works,  some works have also been done in relatively simple scenarios with only a few nodes. 
For example, the authors in \cite{YJu2017TCOM} considered a simple network with only one source equipped with multiple antennas, one destination and one eavesdropper, both having a single antenna. 
The PLS performances of secrecy outage probability and secrecy throughput were evaluated under three beamforming schemes, namely, maximum ratio transmitting (MRT) beamforming, AN beamforming and partial MRT beamforming. 
This work was partially extended in \cite{YJu2019TWCDF}  by  adding a multi-antenna relay. 
In \cite{HZhao2019WCL}, the authors considered the secrecy issue of a secondary transmission pair in a cognitive system, which shares the spectrum of a primary pair while controlling its power to avoid impairing the transmission of the primary pair.  
For PLS performance evaluation, the authors derived the secrecy outage probability of the secondary pair.

\section{System Model and Sight-based Cooperative Jamming Scheme} \label{sec:sys-model}
\subsection{Network Model} 
We consider an mmWave ad hoc network over $\mathbb R^2$, where the locations of  transmitters, \emph{potential} jammers and eavesdroppers are modeled by independent and homogenous PPPs $\Phi_T$, $\Phi_P$ and $\Phi_E$ with densities $\lambda_T$,  $\lambda_P$ and  $\lambda_E$, respectively.  
Each transmitter has a receiver located at a fixed distance $r_0$ away but at a random orientation. 
According to \cite{haenggi2012stochastic}, the locations of the receivers can be modeled by another independent and homogenous PPP $\Phi_R$ with density $\lambda_R=\lambda_T$. 
We assume that the locations of the legitimate nodes (i.e., transmitters, receivers and potential jammers) are known, while those of the eavesdroppers are not. 
We also assume that the eavesdroppers are passive and non-colluding, i.e., decoding messages based on their own observations without sending signals, and the transmitting nodes (i.e., the transmitters and jammers) transmit with the same power $P$.

\subsection{Antenna Model} 
To approximate the antenna patterns of the legitimate nodes, we adopt the sectored antenna model \cite{Bai2015TWC, AThornburg2016TSP}, where each antenna consists of a main lobe and a back lobe. 
We define the beam width of the main lobe, the main lobe gain and back lobe gain of the transmitting nodes' antennas by $\theta_T$, $G_T$ and $g_T$ ($G_T>g_T$), and those of the receivers' antennas by $\theta_R$, $G_R$ and $g_R$ ($G_R>g_R$).
We assume that the transmitter and receiver of each pair have properly steered their antennas to obtain the maximum antenna gain $ G_TG_R$ before transmission. 
Since the PPPs $\Phi_T$, $\Phi_R$ and $\Phi_P$ are isotropic, the random effective antenna gain between a transmitting node  (i.e., a transmitter or jammer) and a receiver is
\begin{align} \label{eqn:ag_d}
\mathsf G =
  \begin{cases}
    G_TG_R,  &  \text{w.p. } q_{G_TG_R}=\frac{\theta_T}{2\pi}\frac{\theta_R}{2\pi}\\
    G_Tg_R,  &  \text{w.p. } q_{G_Tg_R}=\frac{\theta_T}{2\pi}\frac{2\pi - \theta_R}{2\pi} \\
		g_TG_R,  &  \text{w.p. } q_{g_TG_R}=\frac{2\pi-\theta_T}{2\pi}\frac{\theta_R}{2\pi}\\
		g_Tg_R,  &  \text{w.p. } q_{g_Tg_R}=\frac{2\pi-\theta_T}{2\pi}\frac{2\pi-\theta_R}{2\pi}
  \end{cases},
\end{align}
where $q_{\mathsf g}$ ($\mathsf g\in\{G_TG_R, G_Tg_R, g_TG_R, g_Tg_R\}$) denotes the probability of $\mathsf G = \mathsf g$. 

\yet{Similarly, eavesdropper antennas are also characterized by the sectored antenna model, and the random effective antenna gain between an eavesdropper and a transmitting node is
\begin{align} 
\mathsf G_E =
  \begin{cases}
    G_TG_E,  &  \text{w.p. } q_{G_TG_E}=\frac{\theta_T}{2\pi}\frac{\theta_E}{2\pi}\\
    G_Tg_E, &  \text{w.p. } q_{G_Tg_E}=\frac{\theta_T}{2\pi}\frac{2\pi - \theta_E}{2\pi}\\
    g_TG_E,	 &  \text{w.p. } q_{g_TG_E}=\frac{2\pi-\theta_T}{2\pi}\frac{\theta_E}{2\pi}\\
    g_Tg_E,	 &  \text{w.p. } q_{g_Tg_E}=\frac{2\pi-\theta_T}{2\pi}\frac{2\pi - \theta_E}{2\pi}
  \end{cases},
	\label{eqn:ag_E}
\end{align}
where $q_{\hat{\mathsf g}}$ ($\hat{\mathsf g}\in\{G_TG_E, G_Tg_E,g_TG_E,g_Tg_E\}$) denotes the probability of $\mathsf G = \hat{\mathsf g}$,
$G_E$ (resp. $g_E$) denotes the main (resp. back) lobe gain and $\theta_E$ denotes the main beam width of eavesdroppers.}

\subsection{Blockage and Propagation Model}
To depict the blockage effect, we adopt the LoS ball model \cite{WangC2016TWC,AThornburg2016TSP,NDengTWC2017}, where \yet{a link of length $r$ is LoS (resp. NLoS) with probability $p_L$ (resp. $1-p_L$) if $r\leq D$ and with probability $0$ (resp. $1$) otherwise, i.e., if $r>D$}. 
Here, $D$  denotes the radius of LoS balls. 
We assume that $r_0\leq D$ and \emph{the links of the transmission pairs are LoS} throughout this paper. 
The blockage effect results in different path loss for LoS and NLoS links, of which the exponents are denoted by $\alpha_L$ and $\alpha_N$, respectively. 
Besides, the mmWave links also suffer from multi-path fading, which is characterized by the Nakagami-$m$ fading model. 
Thus, the channel gain $h_{x,y}^\mathsf b$ of a link $x\rightarrow y$ follows the gamma distribution $\Gamma(N_{\mathbf b},N_{\mathbf b})$ with shape $N_{\mathbf b}$ and rate $N_{\mathbf b}$, where $\mathbf b = L$ (resp. $\mathbf b=N$) for LoS (resp. NLoS) links.
%Like \cite{NDengTWC2017}, we assume that for a receiving node (i.e., receiver or eavesdropper), \emph{the interference from the interferers outside its LoS ball can be ignored} due to the severe signal attenuation.

\subsection{Sight-based Cooperative Jamming (SCJ) Scheme} \label{sec:SCJ-scheme}
The SCJ scheme aims to select jammers that are expected to generate more interference at the eavesdroppers than at the receivers. 
Since the locations of eavesdroppers are unknown, the potential jammers cannot measure their interference to the eavesdroppers. 
Thus, the selection is solely based on the interference to the receivers. 
To measure such interference, each potential jammer associates itself with the \emph{nearest} receiver and observes the distance $d$ and link $l$ between them. 
For a potential jammer $y\in \Phi_P$, if $d>D$, then $y$ has no LoS link to any receiver and thus will cause little interference to the receivers. 
In this case, $y$ chooses to become a jammer. If $d\leq D$ and $l$ is NLoS, $y$ will cause slight interference to its nearest receiver but may cause severe interference to other receivers when $y$ is inside the LoS ball of more than one receiver. 
To control the interference to other  receivers, $y$ chooses to become a jammer with probability $\rho \in[0,1]$. 
In other cases, $y$ will cause severe interference to the receivers and thus remains silent. 

\yet{Note that the association policy is designed to select the associated receiver for a given potential jammer (rather than selecting associated jammer for a given receiver), with the purpose of reducing as much as possible the interference to other legitimate receivers. 
Thus, a natural association policy is to select the nearest receiver. 
This is because if a potential jammer causes no or little interference to its nearest receiver, it would also cause no or little interference to other receivers as well.
The proposed scheme can be also applied to the mmWave cellular network scenario, since the mmWave cellular networks share some common features with the network scenario considered in this paper. 
For instance, in the uplink transmission scenarios where BSs serve as receivers, idle mobile devices can be the potential jammers. 
By applying the proposed scheme to this scenario, the channel advantages can be created for BSs, leading to an enhancement of STC performance therein.}

We define the region covered by the LoS balls of all the receivers by $\mathcal B = \bigcup_{x\in\Phi_R}B(x,D)$, where $B(x,D)=\{x'\in\mathbb R^2: \lVert x'-x\rVert\leq D\}$ denotes the LoS ball of $x$. 
According to the SCJ scheme, each potential jammer becomes a jammer with probability $\rho p_N$, where $p_N=1-p_L$, if it is inside $\mathcal B$, and with probability $1$ otherwise. 
Thus, the locations of jammers can be equivalently modeled by a homogeneous PPP $\Phi_J$ with density $\lambda_{J}=\rho p_N\lambda_P$ plus an independent Poisson Hole Process (PHP) $\Psi_J$. 
The process $\Psi_J$ is formed by removing the nodes in $\mathcal B$ from another independent and homogenous PPP $\bar{\Phi}_J$ (called baseline PPP) with density $\bar{\lambda}_J=(1-\rho p_N)\lambda_P$. 
Formally, $\Psi_J$ can be defined by $\Psi_J=\bar{\Phi}_J\cap \mathcal B^c=\{y\in \bar{\Phi}_J: y\notin \mathcal B\}$.

\subsection{Performance Metrics}
The transmission pairs apply the Wyner encoding scheme \cite{Wyner1975} to protect their confidential messages. 
When sending a confidential message, a transmitter $x$ chooses a codeword rate $R_s$ for this message and another codeword rate $R_t$ for the entire transmitted message. 
The difference $R_e=R_t-R_s$ reflects the cost for confusing the eavesdroppers. 
We assume $R_t$ and $R_s$ are fixed throughout this paper. 
Due to the random channel condition, the corresponding receiver $y$ succeeds in decoding the transmitted message with a certain probability. This probability is called \textbf{connection probability}  \cite{WangC2016TWC}  and defined by 
\begin{IEEEeqnarray}{rCl}\label{eqn:def-pc}
p_c&=&\mathbb P(\mathrm{SINR}_{x,y}\geq 2^{R_t}-1),
\end{IEEEeqnarray}
where $\mathrm{SINR}_{x,y}$ denotes the Signal-to-Interference-plus-Noise Ratio (SINR) from $x$ to $y$. 
Similarly, eavesdroppers fail to obtain any information from the confidential message with a certain probability. 
This probability is called \textbf{secrecy probability} \cite{WangC2016TWC} and defined by the probability that no eavesdropper can decode the confidential message, that is, 
\begin{IEEEeqnarray}{rCl}\label{eqn:def-ps}
p_s&=&\mathbb P\Big(\bigcap_{z\in \Phi_E} \mathrm{SINR}_{x,z}\leq 2^{R_e}-1\Big),
\end{IEEEeqnarray}
where $\mathrm{SINR}_{x,z}$ denotes the SINR from $x$ to an eavesdropper $z\in \Phi_E$. 
\yet{Note that the concrete expressions of $\mathrm{SINR}_{x,y}$ and $\mathrm{SINR}_{x,z}$ will be determined in the subsequent sections where $x$, $y$ and $z$ are clearly defined.}
%These two probabilities are fundamental and widely-used metrics in the research area, and thus are easier for other researchers to understand and accept. Derivations of these two probabilities are not obvious, as can be seen from the theoretical analysis in subsequent sections.
Using these probabilities, we define another metric, called \textbf{secrecy transmission capacity} (STC) \cite{Zhou2011}, to characterize the average sum rate of the transmissions in perfect secrecy per unit area. Formally, the STC can be formulated as 
\begin{IEEEeqnarray}{rCl} \label{eqn:def-stc}
\bar{R}_s = p_{c}p_{s}(R_t - R_e)\lambda_T.
\end{IEEEeqnarray}
This papers uses the STC $\bar{R}_s$ as the metric to evaluate the secrecy performance, for which we need to derive the connection probability $ p_{c}$ and secrecy probability $p_{s}$ of a typical pair, respectively. 

\section{Performance Analysis: Simplified Scenario}\label{sec:per-ana-simple-case}
To facilitate the STC analysis of the general scenario, we consider a simplified scenario with only one transmission pair and derive the $p_c$ and $p_s$ in Sections \ref{sec:pc-simple-case} and \ref{sec:ps-simple-case}, respectively. 
\yet{The simplified scenario provides an intuitive insight into the main idea of the proposed SCJ scheme and makes it easier for readers to see the superiority of the SCJ scheme in terms of STC performance. 
In addition, by introducing some basic theoretical results in the simplified scenario, we can focus more on the derivations of the key theoretical results in the general scenario, so that the readers can easily understand the main idea of our theoretical analysis without being lost in the cumbersome and tedious mathematical derivations.}

\subsection{Connection Probability}\label{sec:pc-simple-case}
We define $x_0$ and $y_0$ the transmitter and receiver of the pair, respectively. The SINR of $y_0$ can be given by
%\begin{IEEEeqnarray}{rCl}
%\mathrm{SINR}_{x_0, y_0}=\frac{G_TG_Rh_{x_0,y_0}^{L}r_0^{-\alpha_L}}{I_{\Phi_J}^{y_0}+I_{\Psi_J}^{y_0}+\sigma^2/P},
%\end{IEEEeqnarray}
\begin{IEEEeqnarray}{rCl}
\mathrm{SINR}_{x_0, y_0}=\frac{G_TG_Rh_{x_0,y_0}^{L}r_0^{-\alpha_L}}{I_{\Phi_J}^{y_0}+I_{\Psi_J}^{y_0}+\sigma^2/P},
\end{IEEEeqnarray}
where $\sigma^2$ denotes the variance of the noise and $I_{\Phi_J}^{y_0}$ (resp. $I_{\Psi_J}^{y_0}$) is the interference from the jammers in $\Phi_J$ (resp. $\Psi_J$). 
To obtain the connection probability, we first derive the Laplace transform of $I_{\Phi_J}^{y_0}$ and $I_{\Psi_J}^{y_0}$ in the following lemmas.
\begin{lemma}\label{lemma:LT-Phi-J-simple-case}
The Laplace transform $\mathcal L_{I_{\Phi_J}^{y_0}}(s)$ of $I_{\Phi_J}^{y_0}$ in the simplified scenario is 
\begin{IEEEeqnarray}{rCl}\label{eqn:LT-Phi-J-simple-case}
\mathcal L_{I_{\Phi_J}^{y_0}}(s)&=&\exp\Big(\!\!-2\pi\lambda_J\sum_{\mathsf g}q_{\mathsf g} \int_0^\infty F_N(s\mathsf g, r)r\mathrm dr\!\!\Big),
\end{IEEEeqnarray}
where
\begin{align}\label{eqn:Fb}
F_{\mathsf b}(x, r)=1-\Big(1+\frac{x}{N_{\mathsf b}r^{\alpha_{\mathsf b}}}\Big)^{-N_{\mathsf b}},
\end{align}
$\mathsf b\in\{L,N\}$, $\mathsf g\in\{G_TG_R,G_Tg_R, g_TG_R, g_Tg_R\}$ and $q_{\mathsf g}$ is given in (\ref{eqn:ag_d}).
\end{lemma}
\begin{IEEEproof}
See Appendix \ref{app:LT-Phi-J-simple-case}.
\end{IEEEproof}
\begin{lemma}\label{lemma:LT-Psi-J-simple-case}
The Laplace transform $\mathcal L_{I_{\Psi_J}^{y_0}}(s)$ of $I_{\Psi_J}^{y_0}$ in the simplified scenario is 
\begin{IEEEeqnarray}{rCl}\label{eqn:LT-Psi-J-simple-case}
\mathcal L_{I_{\Psi_J}^{y_0}}(s)&=& \exp\Big(\!\!-2\pi\bar{\lambda}_J\sum_{\mathsf g}q_{\mathsf g} \int_D^\infty F_{N}(s\mathsf g, r)r\mathrm dr\!\!\Big),
\end{IEEEeqnarray}
where
\begin{align}\label{eqn:Fb}
F_{N}(x, r)=1-\Big(1+\frac{x}{N_{N}r^{\alpha_{N}}}\Big)^{-N_{N}},
\end{align}
$\mathsf g\in\{G_TG_R,G_Tg_R, g_TG_R, g_Tg_R\}$ and $q_{\mathsf g}$ is given in (\ref{eqn:ag_d}).
\end{lemma}
\begin{IEEEproof}
This proof follows from Lemma \ref{lemma:LT-Phi-J-simple-case}.
\end{IEEEproof}

Based on Lemmas \ref{lemma:LT-Phi-J-simple-case} and \ref{lemma:LT-Psi-J-simple-case}, we now derive the upper bound on the connection probability of the transmission pair. 
\begin{theorem}\label{theorem:pc-simple-case}
The connection probability of the transmission pair in the simplified scenario can be upper bounded by   
\begin{IEEEeqnarray}{rCl}
p_{c}&\leq &\sum_{k=1}^{N_L}\binom{N_L}{k}(-1)^{k+1}e^{-\frac{k\mu \sigma^2}{P}}\mathcal L_{I_{\Phi_J}^{y_0}}(k\mu )\mathcal L_{I_{\Psi_J}^{y_0}}(k\mu ),
\end{IEEEeqnarray} 
where $\mu=\frac{\tau_L r_0^{\alpha_L}(2^{R_t}-1)}{G_TG_R}$, $\tau_L=N_L(N_L!)^{-1/N_L}$, $\mathcal L_{I_{\Phi_J}^{y_0}}(\cdot)$ is given by (\ref{eqn:LT-Phi-J-simple-case}) and $\mathcal L_{I_{\Psi_J}^{y_0}}(\cdot)$ is given by (\ref{eqn:LT-Psi-J-simple-case}).
\end{theorem}
\begin{IEEEproof}
Based on the definition in (\ref{eqn:def-pc}), we have
\begin{IEEEeqnarray}{rCl}
p_{c}&=&\mathbb P\Big(h_{x_0,y_0}^{L}\geq \frac{(2^{R_t}-1) r_0^{\alpha_L}}{G_TG_R}(I_{\Phi_J}^{y_0}+I_{\Psi_J}^{y_0}+\sigma^2/P)\Big)\nonumber\\
&\overset{(a)}\leq&1-\mathbb E_{I_{\Phi_J}^{y_0},I_{\Psi_J}^{y_0}}\Big[\Big(1-e^{-\mu (I_{\Phi_J}^{y_0}+I_{\Psi_J}^{y_0}+\sigma^2/P)}\Big)^{N_L}\Big]\nonumber\\
&=&\sum_{k=1}^{N_L}\binom{N_L}{k}(-1)^{k+1}e^{-k\mu \frac{\sigma^2}{P}}\mathcal L_{I_{\Phi_J}^{y_0}}\left(k\mu\right)\mathcal L_{I_{\Psi_J}^{y_0}}(k\mu),
\end{IEEEeqnarray}
where $(a)$ follows from the Lemma $1$ in \cite{AThornburg2016TSP}.
\end{IEEEproof} 

\subsection{Secrecy Probability }\label{sec:ps-simple-case} 
The PPP $\Phi_E$ inside the LoS ball $B(x_0,D)$ can be divided into independent sub-PPPs $\Phi_{E,I}^{\mathsf b,\hat{\mathsf g}}$ with density $p_{\mathsf b}q_{\hat{\mathsf g}}\lambda_E$ ($\mathsf b\in \{L,N\}$ and $\hat{\mathsf g}\in\{G_TG_E, G_Tg_E,g_TG_E,g_Tg_E\}$) due to the independent thinning by the blockage effect and antenna gains, while the PPP $\Phi_E$ outside $B(x_0,D)$ can be divided into sub-PPPs $\Phi_{E,O}^{N,\hat{\mathsf g}}$ with density $q_{\hat{\mathsf g}}\lambda_E$, because these eavesdroppers have only NLoS links to $x_0$. Since these sub-PPPs are independent, we can formulate the secrecy probability as
\begin{IEEEeqnarray}{rCl}\label{eqn:ps-formulation-simple-case} 
p_s&=&\prod_{\mathsf b}\prod_{\hat{\mathsf g}}\mathbb P\left(\cap_{z\in \Phi_{E,I}^{\mathsf b,\hat{\mathsf g}}} \mathrm{SINR}_{x_0,z}^{\mathsf b,\hat{\mathsf g}}\leq 2^{R_e}-1\right)\\
&&\times\prod_{\hat{\mathsf g}}\mathbb P\left(\cap_{z\in \Phi_{E,O}^{N,\hat{\mathsf g}}} \mathrm{SINR}_{x_0,z}^{N,\hat{\mathsf g}}\leq 2^{R_e}-1\right),\nonumber
\end{IEEEeqnarray} 
where $\mathrm{SINR}_{x_0,z}^{\mathsf b,\hat{\mathsf g}}$ denotes the SINR of an eavesdropper $z$ with link condition $\mathsf b$ (i.e., LoS or NLoS) and antenna gain $\hat{\mathsf g}$ to $x_0$. Here, $\mathrm{SINR}_{x_0,z}^{\mathsf b,\hat{\mathsf g}}$ is given by 
\begin{IEEEeqnarray}{rCl}
\mathrm{SINR}_{x_0,z}^{\mathsf b,\hat{\mathsf g}}=\frac{\hat{\mathsf g} h_{x_0,z}^{\mathsf b}\lVert x_0-z \rVert^{-\alpha_{\mathsf b}}}{\sum_{\varphi\in\{\Phi_J, \Psi_J\}}I_{\varphi}^{z}+\sigma^2/P},
\end{IEEEeqnarray}
where $I_{\varphi}^z$ ($\varphi\in\{\Phi_J, \Psi_J\}$) denotes the interference at $z$ from $\varphi\cap B(z,D)$. For $\varphi=\Phi_J$, $\mathcal L_{I_{\Phi_J}^z}(s)$ can be obtained based on Lemma \ref{lemma:LT-Phi-J-simple-case}. 
We summarize the result in the following lemma.
\begin{lemma}\label{lemma:LT-Phi-J-E-simple-case}
The Laplace transform $\mathcal L_{I_{\Phi_J}^z}(s)$ of $I_{\Phi_J}^z$ in the simplified scenario is 
\begin{IEEEeqnarray}{rCl}\label{eqn:LT-Phi-J-E-simple-case}
\mathcal L_{I_{\Phi_J}^{z}}\!\!(\lambda_J, s)\!&=&\!\exp\Big(\!-\!2\pi\lambda_J\!\!\sum_{\mathsf b}\!\!\sum_{\hat{\mathsf g}}p_{\mathsf b}q_{\hat{\mathsf g}}\!\! \int_0^D\!\! F_{\mathsf b}(s\hat{\mathsf g}, r)r\mathrm dr\!\Big)\nonumber\\
&&\times \exp\Big(\!-\!2\pi\lambda_J\!\!\sum_{\hat{\mathsf g}}q_{\hat{\mathsf g}}\!\! \int_D^\infty\!\! F_{N}(s\hat{\mathsf g}, r)r\mathrm dr\!\Big),
\end{IEEEeqnarray}
where $F_{\mathsf b}(\cdot,\cdot)$ is given by (\ref{eqn:Fb}).
\end{lemma}
\begin{IEEEproof}
The proof is similar to that of Lemma \ref{lemma:LT-Phi-J-simple-case} and thus omitted here.
\end{IEEEproof}

\begin{figure}[h]
 \centering
 \includegraphics[width=0.4\textwidth]{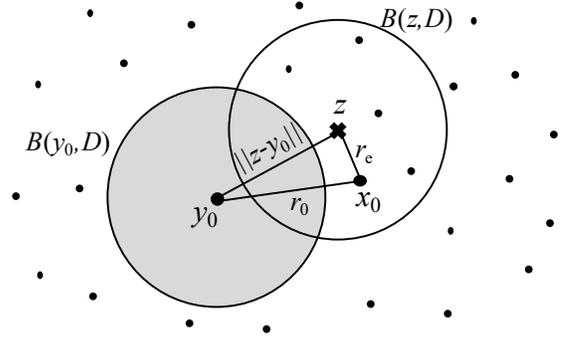}
 \caption{Illustration of jammers inside $B(z,D)$ removed by $B(y_0,D)$.} 
 \label{fig:hole-effect}
\end{figure}

Next, we derive the Laplace transform $\mathcal L_{I_{\Psi_J}^z}(s)$ of $I_{\Psi_J}^z$. 
According to the SCJ scheme, some jammers inside the LoS ball $B(z,D)$ of the eavesdropper $z$ will be removed by the LoS ball $B(y_0,D)$ of the receiver $y_0$ (as shown in Fig.~\ref{fig:hole-effect}), if the distance $\lVert z-y_0\rVert$ between $z$ and $y_0$ is smaller than $2D$. 
Thus, $\mathcal L_{I_{\Psi_J}^z}(s)$ depends on $\lVert z-y_0\rVert$ and thus the distance $\lVert z-x_0\rVert$ between $x_0$ and $z$, because $\lVert z-y_0\rVert$ can be expressed by $\lVert z-x_0\rVert$ and the link distance $r_0$ between $x_0$ and $y_0$. 
Denoting $\lVert z-x_0\rVert$ as $r_e$, we rewrite $\mathcal L_{I_{\Psi_J}^z}(s)$ as $\mathcal L_{I_{\Psi_J}^z}(s, r_e)$ and derive its expression in the following lemma. 
\begin{lemma}\label{lemma:LT-Psi-J-E-simple-case}
The Laplace transform $\mathcal L_{I_{\Psi_J}^z}(s, r_e)$ of $I_{\Psi_J}^z$ in the simplified scenario is
\begin{IEEEeqnarray}{rCl}\label{eqn:LT-Psi-J-E-simple-case}
\mathcal L_{I_{\Psi_J}^z}\!\!(s,r_e)\! &=&\int_{\lvert r_e\!-\!r_0\rvert}^{r_e\!+\!r_0}e^{- \bar{\lambda}_J\sum_{\mathsf b}\sum_{\mathsf g}p_{\mathsf b}q_{\hat{\mathsf g}}T(s,u)}\nonumber\\
&&\	\	e^{-\widetilde{\lambda}_J\sum_{\hat{\mathsf g}}q_{\hat{\mathsf g}} T_2\left(s,u\right)}h(u)\mathrm du,
\end{IEEEeqnarray}
where  
\begin{IEEEeqnarray}{rCl}\label{eqn:T_1}
T_1(s,u)&=&2\pi \int_{C_1(u)}^{D}\!\!\! F_{\mathsf b}(s\hat{\mathsf g},r)r\mathrm d r\\
&&-2\int_{C_2(u)}^{D}\!\!\!F_{\mathsf b}(s\hat{\mathsf g},r)\arccos\left(\frac{u^2\!+\!r^2\!-\!D^2}{2ur}\right)r \mathrm d r \nonumber,
\end{IEEEeqnarray}
\begin{IEEEeqnarray}{rCl}\label{eqn:T_2}
T_2(s,u)&=&2\pi\int_{D}^{\infty}F_{N}(s\hat{\mathsf g},r)r\mathrm d r\\
&&-2\int_{C_3(u)}^{u+D}F_{N}(s\hat{\mathsf g},r)\arccos\left(\frac{u^2+r^2-D^2}{2u r}\right)r\mathrm d r,\nonumber
\end{IEEEeqnarray}
$C_1(u)=\min\{D,\max\{0,D\!-\!u\}\}$, $C_2(u)=\min\{D, \max\{u-D,D-u\}\}$, $C_3(u)=\max\{D,u-D\}$ and 
\begin{align}\label{eqn:pdf-h}
h(u)=\frac{2u}{\pi\sqrt{4r_0^2r_e^2 - (r_0^2+r_e^2-u^2)^2}}.
\end{align}
\end{lemma}
\begin{IEEEproof}
See Appendix \ref{app:LT-Psi-J-E-simple-case}.
\end{IEEEproof}

Based on Lemmas \ref{lemma:LT-Phi-J-E-simple-case} and \ref{lemma:LT-Psi-J-E-simple-case}, we derive the lower bound on the secrecy probability in the following theorem. 
\begin{theorem}\label{theorem:ps-simple-case}
The secrecy probability $p_s$ of the transmission pair in the simplified scenario can be lower bounded by  
\begin{IEEEeqnarray}{rCl}
&p_s&\geq\exp\Bigg(-2\pi \lambda_E \sum_{\mathsf b}\sum_{\hat{\mathsf g}} p_{\mathsf b}q_{\hat{\mathsf g}} \sum_{k=1}^{N_{\mathsf b}}\binom{N_{\mathsf b}}{k}(-1)^{k+1}\\
&&\int_{0}^D e^{-k\nu_{\mathsf b}r_e^{\alpha_{\mathsf b}}\frac{\sigma^2}{ P}}\mathcal L_{I_{\Phi_J}^z}( \lambda_J, k \nu_{\mathsf b} r_e^{\alpha_{\mathsf b}})\mathcal L_{I_{\Psi_J}^z}( k \nu_{\mathsf b} r_e^{\alpha_{\mathsf b}}, r_e)r_e\mathrm d r_e\Bigg)\nonumber\\
&&\times\exp\Bigg(-2\pi \lambda_E\sum_{\hat{\mathsf g}}q_{\hat{\mathsf g}} \sum_{k=1}^{N_N}\binom{N_N}{k}(-1)^{k+1}\int_{D}^\infty  \nonumber\\
&&e^{-k\nu_N r_e^{\alpha_N}\frac{\sigma^2}{ P}}\mathcal L_{I_{\Phi_J}^z}(\lambda_J, k \nu_N r_e^{\alpha_N})\mathcal L_{I_{\Psi_J}^z}(k \nu_N r_e^{\alpha_N}, r_e)r_e\mathrm d r_e\Bigg)\nonumber,
\end{IEEEeqnarray}
where $\nu_{\mathsf b} = \frac{\tau_{\mathsf b}(2^{R_e}-1)}{\hat{\mathsf g}}$, $\tau_{\mathsf b}=N_{\mathsf b}(N_{\mathsf b}!)^{-1/N_{\mathsf b}}$, $\mathcal L_{I_{\Phi_J}^z}(\cdot)$ is given by Lemma \ref{lemma:LT-Phi-J-E-simple-case} and $\mathcal L_{I_{\Psi_J}^z}(\cdot, \cdot)$ by Lemma \ref{lemma:LT-Psi-J-E-simple-case}.
\end{theorem}
\begin{IEEEproof}
See Appendix \ref{app:theorem-ps-simple-case}.
\end{IEEEproof}

 \begin{table}[h]
\renewcommand{\arraystretch}{1.5}
\caption{Parameters used in simulations.}
\label{tb:pmts}
\centering
\begin{tabular}{|l|l|}
\hline
\bfseries Parameters &  \bfseries Value \\
\hline
Link distance $r_0$ & $100$ m\\
Channel bandwidth & $1$ GHz \\
Noise spectral density & $-174$ dBm/Hz\\
Common transmit power $P$ & $1$ W (i.e., $30$ dBm)\\
Path loss exponent $\alpha_L$ ($\alpha_N$) & 2 (4)\\
Nakagami fading parameter $N_L$ ($N_N$) & 3 (2)\\
LoS probability $p_L$ & $0.2$\\
LoS ball radius $D$ & $200$ m\\
Main lobe beam width $\theta_T$ ($\theta_R$)&$\pi/6$ ($\pi/6$)\\
Main lobe gain $G_T$ ($G_R$, \yet{$G_E$}) & $10$ ($10$, \yet{$10$})\\
Back lobe gain $g_T$ ($g_R$, \yet{$g_E$}) & $0.1$ ($0.1$, \yet{$0.1$})\\
\hline
\end{tabular}
\end{table}

\begin{figure}[h]
 \centering
 \includegraphics[width=0.45\textwidth]{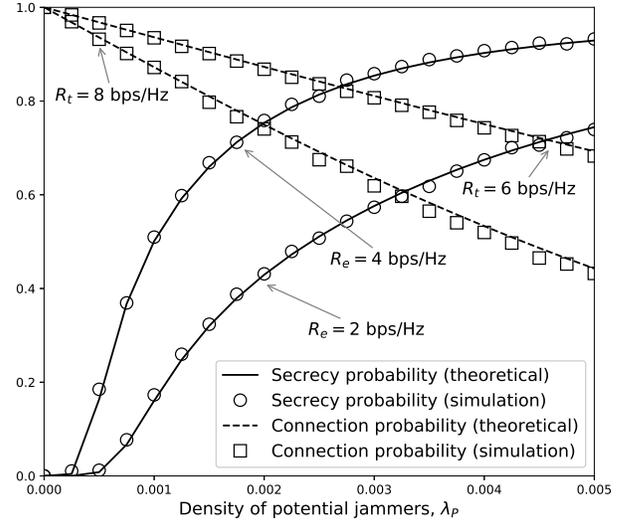}
 \caption{Validation of the bounds in Theorems \ref{theorem:pc-simple-case} and \ref{theorem:ps-simple-case} with $\rho=1.0$ and $\lambda_E=0.005$ based on Monte Carlo simulation. Simulation setting: a circular network with radius $500$ m. Each simulation value is obtained as the arithmetic mean of $10000$ simulation outcomes.}
 \label{fig:validation-simple}
\end{figure}

The bounds in Theorems \ref{theorem:pc-simple-case} and \ref{theorem:ps-simple-case} are validated by Monte Carlo simulations based on the simulator in  \cite{Simu2019TIFS} under different values of $\lambda_P$, $\lambda_E$, $R_t$ and $R_e$.  
\yet{For simulation, we consider an urban environment and summarize in Table \ref{tb:pmts} the related environment parameters. 
Here, the setting of $D=200$ [m] corresponds to the urban area of Manhattan \cite{Singh2015JSAC}, which provides good fit for real-world scenarios.}
The simulation results as well as the theoretical ones are shown in Fig.~\ref{fig:validation-simple}, from which we can see that the theoretical results match nicely with the simulation ones, implying that \emph{the bounds are tight enough to be used as approximations}. 

\section{Performance Analysis: General Scenario}\label{sec:per-ana-general-case}
This section derives the connection and secrecy probabilities of the general scenario with multiple transmission pairs. We focus on a typical  pair and again define the transmitter and receiver by $x_0$ and $y_0$, respectively. 
\subsection{Connection Probability}\label{sec:pc-general-case}
In this scenario, $y_0$ receives interference  from not only the jammers in $\Phi_J$ and $\Psi_J$ but also the concurrent transmitters in $\Phi_T\cap B(y_0,D)$. Thus, the SINR of $y_0$ is given by 
\begin{IEEEeqnarray}{rCl}
\mathrm{SINR}_{x_0, y_0}=\frac{G_TG_Rh_{x_0,y_0}^{L}r_0^{-\alpha_L}}{\sum_{\varphi\in\{\Phi_T, \Phi_J,\Psi_J\}}I_{\varphi}^{y_0}+\sigma^2/P},
\end{IEEEeqnarray}
where $I_{\varphi}^{y_0}$ ($\varphi\in\{\Phi_T, \Phi_J\}$) denotes the interference from $\varphi\cap B(y_0,D)$. Note that  the Laplace transform of $I_{\Phi_T}^{y_0}$ can be easily obtained based on Lemma \ref{lemma:LT-Phi-J-E-simple-case}, which is given in the following lemma.
\begin{lemma}\label{lemma:LT-T-general-case}
The Laplace transform of $I_{\Phi_T}^{y_0}$ is 
\begin{IEEEeqnarray}{rCl}\label{eqn:LT-T-general-case}
\mathcal L_{I_{\Phi_T}^{y_0}}(\lambda_T, s)\!&=&\!\exp\Big(\!\!-\!2\pi \lambda_T \!\sum_{\mathsf b}\!\sum_{\mathsf g}p_{\mathsf b}q_{\mathsf g}\! \!\int_0^D\!\! F_{\mathsf b}(s\mathsf g, r)r\mathrm dr\!\!\Big)\nonumber\\
&&\times \exp\Big(\!\!-\!2\pi \lambda_T \!\sum_{\mathsf g}q_{\mathsf g}\! \!\int_D^\infty\!\! F_{N}(s\mathsf g, r)r\mathrm dr\!\!\Big),
\end{IEEEeqnarray}
where $F_{\mathsf b}(\cdot,\cdot)$ is given by (\ref{eqn:Fb}).
\end{lemma}
\begin{IEEEproof}
The proof follows directly from Lemma \ref{lemma:LT-Phi-J-E-simple-case}.
\end{IEEEproof}

\begin{figure}[h]
 \centering
 \includegraphics[width=0.4\textwidth]{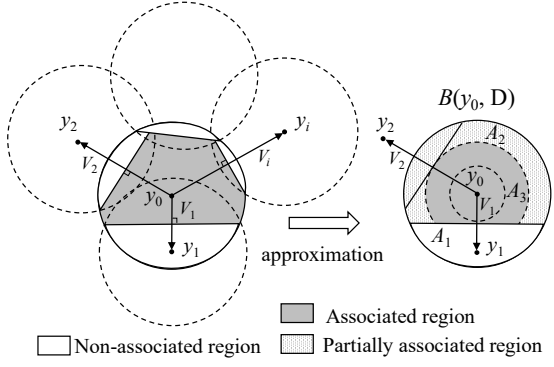}
 \caption{Approximation of associated region. $y_i$ ($i=1,2,\cdots$): the $i$-th nearest receiver to $y_0$; $V_i$: random distance between $y_i$ and $y_0$.}
 \label{fig:region-approx}
\end{figure}

Next, we derive the Laplace transform of $I_{\Phi_J}^{y_0}$. 
According to the SCJ scheme, the LoS ball $B(y_0,D)$ of $y_0$ may contain jammers that are not associated with $y_0$, i.e., jammers that are closer to other receivers than $y_0$. We call these jammers non-associated jammers and the region where they reside non-associated region. 
Fig.~\ref{fig:region-approx} shows an example of the non-associated region, where $y_i$ denotes the $i-$th closest  receiver to $y_0$ and $V_i$ denotes the corresponding random distance. 
As shown in Fig.~\ref{fig:region-approx}, the shape of the non-associated region is irregular, making it difficult to derive the exact Laplace transform of $I_{\Phi_J}^{y_0}$. 
Thus, we resort to a good approximation, which is given in the following lemma.
\begin{lemma}\label{lemma:LT-Phi-J-general-case}
The Laplace transform of $I_{\Phi_J}^{y_0}$ in the general scenario can be approximated by 
\begin{IEEEeqnarray}{rCl}\label{eqn:LT-Phi-J-general-case}
\mathcal L_{I_{\Phi_J}^{y_0}}(s)&\approx& \exp\Big(\!\!-2\pi \lambda_J\sum_{\mathsf g}q_{\mathsf g} \int_0^\infty F_N(s\mathsf g, r)r\mathrm dr\!\!\Big)\\
&&\times\int_{0}^{\infty}\int_{v_1}^{\infty}\exp\Big(-2 p_L\lambda_{J}\sum_{\mathsf g}q_{\mathsf g}Q_1(s,v_1) \Big)\nonumber\\
&&\ \ \ \ \ \ \ \ \exp\Big(-2 \xi(v_2)p_L\lambda_{J}\sum_{\mathsf g}q_{\mathsf g}Q_2(s,v_1,v_2)\Big)\nonumber\\
&&\ \ \ \ \ \ \ \ \ f_{V_1,V_2}(v_1,v_2)\mathrm dv_2 \mathrm dv_1, \nonumber
\end{IEEEeqnarray}
where
\begin{IEEEeqnarray}{rCl}\label{eqn:Q1}
Q_1(s,\! v_1)\!&=&\!\int_{\min\{\frac{v_1}{2},D\}}^{D}\!\!\Big(F_L(s\mathsf g, r)\!-\! F_N(s\mathsf g, r)\Big)\\
&&\ \ \ \ \ \  \ \ \ \ \ \ \ \ \arccos\!\!\left(\frac{v_1}{2r}\right) r\mathrm dr, \nonumber
\end{IEEEeqnarray}
\begin{IEEEeqnarray}{rCl}\label{eqn:Q2}
Q_2(s,\!v_1,\!v_2)\!&=&\!\int_{\min\{\frac{v_2}{2},D\}}^{D}\!\!\Big(F_L(s\mathsf g, r)\!-\! F_N(s\mathsf g, r)\Big)\\
&&\ \ \ \ \ \ \ \ \ \ \left(\pi - \arccos\!\!\left(\frac{v_1}{2r}\right)\right) r\mathrm dr, \nonumber
\end{IEEEeqnarray}
\begin{align}\label{eqn:varrho}
\xi(v_2)=\min\Big\{\frac{2\lambda_R\int_{v_2}^{2D}A(r)r\mathrm d r}{D^2-(v_2/2)^2},1\Big\},
\end{align}
\begin{align}\label{eqn:A}
A(r)=D^2\arccos\Big(\frac{r}{2D}\Big)-\frac{r}{2}\sqrt{D^2-\Big(\frac{r}{2}\Big)^2},
\end{align}
\begin{align}\label{eqn:pdf-V1V2}
f_{V_1,V_2}(v_1,v_2)=(2\pi \lambda_R)^2 v_1v_2 \exp\left(- \lambda_R\pi v_2^2\right).
\end{align}
\end{lemma}
\begin{IEEEproof}
The idea is partitioning the LoS ball $B(y_0,D)$ into three parts as illustrated in Fig.~\ref{fig:region-approx}, i.e., the \emph{non-associated region} $A_1$, where the jammers are associated with $y_1$, the ring-like \emph{partially associated region} $A_2$, where part of the jammers are associated with the remaining receivers $y_i$ ($i=2,3,\cdots$), and the \emph{associated region} $A_3$, where the jammers are associated with $y_0$. 
According to the SCJ scheme, the locations of the jammers in $A_1$ and $A_3$ can be modeled by homogeneous PPPs. 
The most challenging task is to model the locations of the jammers in $A_2$, since each receiver $y_i$ ($i=2,3,\cdots$) carves out a sub-non-associated region from $A_2$ and these sub-non-associated regions may overlap, forming an irregular non-associated region (see Fig.~\ref{fig:region-approx}). 
To tackle this challenge, we remove the non-associated region from $A_2$ and approximate the locations of LoS and NLoS jammers in $A_2$ by two independent and homogenous PPPs with densities $\xi p_L\lambda_J$ and $(1-\xi p_L)\lambda_J$ respectively. 
Here, $\xi\in[0,1]$ denotes the probability that a jammer in $A_2$ is located in the non-associated region. For the detailed proof, please refer to Appendix \ref{app:LT-Phi-J-general-case}. 
\end{IEEEproof}

Next, we derive the Laplace transform of $I_{\Psi_J}^{y_0}$.
Note that $\Psi_J$ is a Poisson Hole Process (PHP), making it challenging to derive the exact $\mathcal L_{I_{\Psi_J}^{y_0}}(s)$ due to the difficulty of accurately characterizing the impacts of the holes, i.e., the LoS balls of the receivers. 
One approach is to consider only the hole $B(y_0,D)$ and neglect all the other holes.
This is equivalent to the simplified scenario. 
Since the interferers in $\Psi_J$ have only NLoS links to $y_0$ and are far away, this approach yields accurate approximation.
Based on this  approach, the Laplace transform of $I_{\Psi_J}^{y_0}$ in the generalized scenario can be approximated by that of the simplified scenario in Lemma \ref{lemma:LT-Psi-J-simple-case}.
Based on Lemmas \ref{lemma:LT-Psi-J-simple-case}, \ref{lemma:LT-T-general-case} and \ref{lemma:LT-Phi-J-general-case}, we derive the connection probability in the following theorem.
\begin{theorem}\label{theorem:pc-general-case}
The connection probability of the transmission pairs in the general scenario is approximated by   
\begin{IEEEeqnarray}{rCl}
p_{c}\!&\approx&\!\sum_{k=1}^{N_L}\!\binom{N_L}{k}\!(-1)^{k+1}\!e^{-\frac{k\mu \sigma^2}{P}}\nonumber\\
&&\ \ \ \ \ \ \mathcal L_{I_{\Phi_T}^{y_0}}(\lambda_T, k\mu)\mathcal L_{I_{\Phi_J}^{y_0}}(k\mu)\mathcal L_{I_{\Psi_J}^{y_0}}(k\mu),
\end{IEEEeqnarray} 
where $\mu=\frac{\tau_L  r_0^{\alpha_L}(2^{R_t}-1)}{G_TG_R}$, $\tau_L=N_L(N_L!)^{-1/N_L}$, $\mathcal L_{I_{\Phi_T}^{y_0}}(\cdot)$ is given by (\ref{eqn:LT-T-general-case}), $\mathcal L_{I_{\Phi_J}^{y_0}}(\cdot)$ is given by (\ref{eqn:LT-Phi-J-general-case}) and $\mathcal L_{I_{\Psi_J}^{y_0}}(\cdot)$ is given by (\ref{eqn:LT-Psi-J-simple-case}).
\end{theorem}
\begin{IEEEproof}
The proof follows after replacing the inequality in Theorem \ref{theorem:pc-simple-case} with approximation.
\end{IEEEproof}

\begin{figure}[h]
 \centering
 \includegraphics[width=0.45\textwidth]{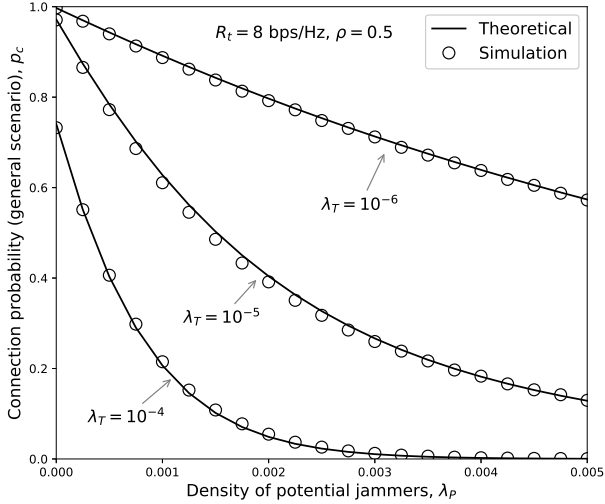}
 \caption{Validation of the approximation in Theorem \ref{theorem:pc-general-case} based on Monte Carlo simulation. Simulation setting: a circular network with radius $500$m. Each simulation value is obtained as the arithmetic mean of $10000$ simulation outcomes.}
 \label{fig:validation-pco-general}
\end{figure}

Using the simulator in \cite{Simu2019TIFS}, we also conducted simulations for the connection probability under various settings of $\lambda_T$ and $\lambda_P$ to verify the effectiveness of the approximation in Theorem \ref{theorem:pc-general-case}. 
The simulation results as well as the theoretical ones are shown in Fig.~\ref{fig:validation-pco-general}. 
These results indicate that the approximation is accurate and the analytical expression in Theorem \ref{theorem:pc-general-case} is effective to model the connection probability of the transmission pairs in the general case.

For comparison, we also investigate the connection probabilities of the case with partial jamming (PJ), where, according to the jamming scheme in \cite{YZhu2018JSAC}, a fraction $\varrho$ of  the potential jammers send artificial noise. 
This can be interpreted as the case where each potential jammer decides to be a jammer independently with probability $\varrho$. 
Thus, the resulting PPP of the jammers is simply an independent thinning of the original PPP $\Phi_P$ of the potential jammers. 
Based on Theorem \ref{theorem:pc-general-case}, the connection probability of the PJ scheme can be given by the following corollary.
\begin{corollary}
The connection probability of the transmission pairs under the PJ scheme can be given by 
\begin{IEEEeqnarray}{rCl}
p_{c}^{\mathrm{PJ}}\!&\approx&\!\sum_{k=1}^{N_L}\!\binom{N_L}{k}(-1)^{k+1}e^{-\frac{k\mu \sigma^2}{P}}\mathcal L_{I_{\Phi_T}^{y_0}}(\lambda_T\!+\! \varrho \lambda_P, k\mu),
\end{IEEEeqnarray}
where  $\mu=\frac{\tau_L  r_0^{\alpha_L}(2^{R_t}-1)}{G_TG_R}$, $\tau_L=N_L(N_L!)^{-1/N_L}$ and $\mathcal L_{I_{\Phi_T}^{y_0}}(\cdot, \cdot)$ is given by (\ref{eqn:LT-T-general-case}).
\begin{proof}
The proof follows directly from Theorem \ref{theorem:pc-general-case}.
\end{proof} 
\end{corollary}

\subsection{Secrecy Probability}
In the general scenario, the SINR of an eavesdropper $z$ is 
\begin{IEEEeqnarray}{rCl}
\mathrm{SINR}_{x_0,z}^{\mathsf b,\hat{\mathsf g}}=\frac{\hat{\mathsf g} h_{x_0,z}^{\mathsf b}\lVert x_0-z \rVert^{-\alpha_{\mathsf b}}}{\sum_{\varphi\in\{\Phi_T, \Phi_J, \Psi_J\}}I_{\varphi}^{z}+\sigma^2/P},
\end{IEEEeqnarray}
where $I_{\varphi}^z$ ($\varphi\in\{\Phi_T, \Phi_J, \Psi_J\}$) denotes the interference from the interferers in $\varphi\cap B(z,D)$.
Note that the Laplace transform of $I_{\Phi_T}^z$ can be directly given based on that of $I_{\Phi_J}^z$ in Lemma \ref{lemma:LT-Phi-J-E-simple-case}. Thus, we have 
\begin{IEEEeqnarray}{rCl}\label{eqn:LT-Phi-T-E-general-case}
\mathcal L_{I_{\Phi_T}^{z}}(\lambda_T, s)=\mathcal L_{I_{\Phi_J}^{z}}(\lambda_T, s),
\end{IEEEeqnarray}
where $\mathcal L_{I_{\Phi_J}^{z}}(\cdot, \cdot)$ is given by Lemma \ref{lemma:LT-Phi-J-E-simple-case}.

Next, we derive the Laplace transform $\mathcal L_{I_{\Psi_J}^z}(s)$ of $I_{\Psi_J}^z$, in which we need to deal with the PHP $\Psi_J$.
According to the definition of $\Psi_J$, each node of the baseline PPP $\bar{\Phi}_J$ is retained (i.e., not removed by the holes) with probability $e^{-\lambda_R \pi D^2}$. 
Thus, one approach to characterize the impact of holes is to approximate $\Psi_J$ by a homogeneous PPP with density $e^{-\lambda_R \pi D^2}\bar{\lambda}_J$. 
\yet{This approach (say Approach 1) reduces the density of interferers in the neighborhood of $z$, and may underestimate the interference to some extent and result in an upper bound on the Laplace transform}, as shown by the figures (Fig. $7$ - Fig. $12$) in \cite{YazdanshenasanTWC2016}. 
Another approach is to neglect the holes, i.e., approximate $\Psi_J$ by a homogeneous PPP with density $\bar{\lambda}_J$. 
This approach overestimates the interference and thus yields a lower bound. 
To further improve this lower bound, the authors in \cite{YazdanshenasanTWC2016} considered only the hole closest to $z$ and showed that the new bound is tight enough. 
However, the typical point $z$ in \cite{YazdanshenasanTWC2016} is assumed to be located outside the holes. 
As a result, this new approach \yet{(say Approach 2)} cannot be directly applied in our case, where $z$ can be inside or outside the holes. 
This paper therefore combines the above three approaches to approximate $\mathcal L_{I_{\Psi_J}^z}(s)$. 
Specifically, we first approximate $\Psi_J$ by a homogeneous $\widetilde{\Phi}_J$ with density 
\begin{align}\label{eqn:lambda_widetilde}
\widetilde{\lambda}_J= (\beta e^{-\lambda_R \pi D^2}+(1-\beta))\bar{\lambda}_J,
\end{align}
where $\beta \in[0,1]$, and then carve out the hole closest to $z$ from $\widetilde{\Phi}_J$. 
The parameter $\beta$ here is used to adjust the accuracy of the approximation. 
The approximation of $\mathcal L_{I_{\Psi}^z}(s)$ is given in the following lemma.  
\begin{lemma}\label{lemma:LT-Psi-J-E-general-case}
The Laplace transform $\mathcal L_{I_{\Psi}^z}(s)$ of the general scenario can be approximated by 
\begin{IEEEeqnarray}{rCl}
\mathcal L_{I_{\Psi_J}^z}(s,r_e) &\approx& \int_{0}^{\infty}\int_{\lvert r_e-r_0\rvert}^{r_e+r_0}e^{-\widetilde{\lambda}_J\sum_{\mathsf b}\sum_{\hat{\mathsf g}}p_{\mathsf b}q_{\hat{\mathsf g}} T_1\left(s,\min\{u,v\}\right)}\nonumber\\
&&e^{-\widetilde{\lambda}_J\sum_{\hat{\mathsf g}}q_{\hat{\mathsf g}} T_2\left(s,\min\{u,v\}\right)} h(u)f(v)\mathrm du \mathrm dv, 
\end{IEEEeqnarray} 
where $\widetilde{\lambda}_J$ is given by (\ref{eqn:lambda_widetilde}), $T_1(\cdot, \cdot)$ by (\ref{eqn:T_1}),  $T_2(\cdot, \cdot)$ by (\ref{eqn:T_2}), $h(u)$ by (\ref{eqn:pdf-h}) and $f(v)=2 \pi \lambda_R v e^{-\lambda_R \pi v^2}$.
\end{lemma}
\begin{IEEEproof}
We define the receiver closest to $z$ by $\tilde{y}$. Note that $\tilde{y}$ can be $y_0$ or the receiver closest to $z$ except $y_0$ (say $y^*$). 
Thus, assuming $z$ is at the origin, we have $\lVert \tilde{y}\rVert=\min\{\lVert y_0\rVert, \lVert y^*\rVert\}$. 
It follows from Appendix \ref{app:LT-Psi-J-E-simple-case} that 
\begin{IEEEeqnarray}{rCl}\label{eqn:LT-Psi-J-E-general-case-2}
\mathcal L_{I_{\Psi_J}^z}(s,r_e)\!&\approx&\!\mathbb E_{\lVert \tilde{y}\rVert}\Big[e^{-\widetilde{\lambda}_J\sum_{\mathsf b }\sum_{\hat{\mathsf g} }p_{\mathsf b}q_{\hat{\mathsf g}}T(s,\lVert \tilde{y}\rVert)}\nonumber\\
&&\ \ \ \ \ \ e^{-\widetilde{\lambda}_J\sum_{\hat{\mathsf g}}q_{\hat{\mathsf g}} T_2\left(s,\lVert \tilde{y}\rVert\right)} \Big].
\end{IEEEeqnarray} 
The PDF of $\lVert y_0\rVert$ can be given by (\ref{eqn:pdf-h}) and that of $\lVert y^*\rVert$ by $f(v)$ according to \cite{Chiu2013}. Substituting $\lVert \tilde{y}\rVert=\min\{\lVert y_0\rVert, \lVert y^*\rVert\}$ into (\ref{eqn:LT-Psi-J-E-general-case-2}) and then taking the expectation in terms of $\lVert y_0\rVert$ and  $\lVert y^*\rVert$ completes the proof.
\end{IEEEproof}

\yet{Please note that our analysis can actually cover the two approaches in \cite{YazdanshenasanTWC2016} as special cases. 
For example, the Laplace transform result of Approach 1 can be obtained by Lemma \ref{lemma:LT-Phi-J-E-simple-case} when setting $\lambda_J=e^{-\lambda_R \pi D^2} \bar{\lambda}_J$ and the Laplace transform result of Approach 2 can be obtained by Lemma \ref{lemma:LT-Psi-J-E-general-case} when setting $\beta=0$.}
Based on Lemmas \ref{lemma:LT-Phi-J-E-simple-case} and \ref{lemma:LT-Psi-J-E-general-case}, we now derive the secrecy probability of the transmission pairs for the general scenario.
\begin{theorem}\label{theorem:ps-general-case}
The secrecy probability $p_s$ of the transmission pairs in the general scenario can be approximated by  
\begin{IEEEeqnarray}{rCl}
&&p_s\approx\exp\Bigg(\!\!-2\pi \sum_{\mathsf b}\sum_{\hat{\mathsf g}}p_{\mathsf b}q_{\hat{\mathsf g}}\lambda_E \!\!\sum_{k=1}^{N_{\mathsf b}}\binom{N_{\mathsf b}}{k}(-1)^{k+1}\!\!\int_{0}^D\!\!\mathrm d r_e\nonumber\\
&&\ \ \  e^{-k\nu_{\mathsf b} r_e^{\alpha_{\mathsf b}}\frac{\sigma^2}{ P}}\mathcal L_{I_{\Phi_J}^z}( \lambda_T+\lambda_J, k \nu_{\mathsf b} r_e^{\alpha_{\mathsf b}})\mathcal L_{I_{\Psi_J}^z}\!\!( k \nu_{\mathsf b} r_e^{\alpha_{\mathsf b}}, r_e)r_e\!\!\Bigg)\nonumber\\
&&\times\exp\Bigg(\!\!-2\pi\sum_{\hat{\mathsf g}} q_{\hat{\mathsf g}}\lambda_E \sum_{k=1}^{N_{N}}\binom{N_{N}}{k}(-1)^{k+1}\!\!\int_{D}^\infty\!\! e^{-k\nu_{N} r_e^{\alpha_{N}}\frac{\sigma^2}{ P}}\nonumber\\
&&\ \ \mathcal L_{I_{\Phi_J}^z}\!\!(\lambda_T+\lambda_J, k \nu_{N} r_e^{\alpha_{N}})\mathcal L_{I_{\Psi_J}^z}\!\!( k \nu_{N} r_e^{\alpha_{N}}, r_e)r_e\mathrm d r_e\!\!\Bigg) , 
\end{IEEEeqnarray}
where $\nu_{\mathsf b} = \frac{\tau_{\mathsf b}(2^{R_e}-1)}{\hat{\mathsf g}}$, $\tau_{\mathsf b}=N_{\mathsf b}(N_{\mathsf b}!)^{-1/N_{\mathsf b}}$, $\mathcal L_{I_{\Phi_J}^z}(\cdot, \cdot)$ and $\mathcal L_{I_{\Psi_J}^z}(\cdot, \cdot)$ are given by Lemmas  \ref{lemma:LT-Phi-J-E-simple-case} and \ref{lemma:LT-Psi-J-E-general-case}, respectively.
\end{theorem}
\begin{IEEEproof}
The proof follows after replacing the inequality in Theorem \ref{theorem:ps-simple-case} by approximation.
\end{IEEEproof}

\begin{figure}[h]
 \centering
 \includegraphics[width=0.45\textwidth]{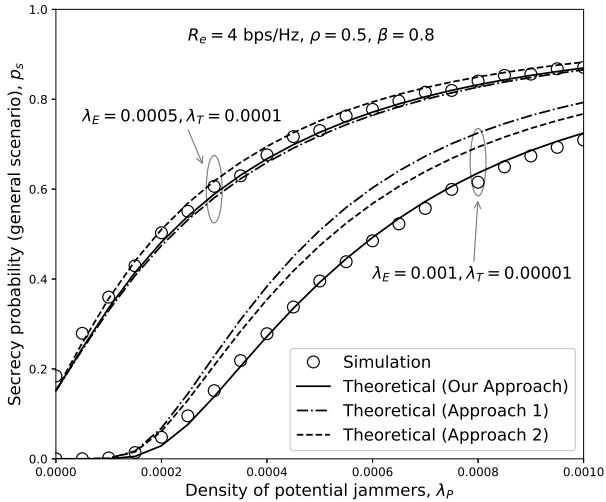}
 \caption{Validation of the approximation in Theorem \ref{theorem:ps-general-case} based on Monte Carlo simulation. Simulation setting: a circular network with radius $500$m. Each simulation value is obtained as the arithmetic mean of $100000$ simulation outcomes.}
 \label{fig:validation-psec-general}
\end{figure}

Simulations based on the simulator in \cite{Simu2019TIFS} were also conducted under various settings of $\lambda_T$, $\lambda_E$ and $\lambda_P$ to demonstrate the effectiveness of the approximated secrecy probability in Theorem \ref{theorem:ps-general-case}. 
The simulation results as well as the corresponding theoretical ones obtained by our approach and the two approaches in \cite{YazdanshenasanTWC2016} are shown in Fig.~\ref{fig:validation-psec-general}. 
These results indicate that the approximation in Theorem \ref{theorem:ps-general-case} is effective to model the secrecy probability of the transmission pairs in the general scenario. 
\yet{In addition, the results show that our approach gives more accurate approximation to secrecy probabilities than the above two approaches in \cite{YazdanshenasanTWC2016}, especially for small densities of transmitters.} 

We now justify  the superiority of the proposed SCJ scheme in terms of STC performance enhancement.
 \begin{lemma}
\yet{The proposed SCJ scheme improves the network STC under the general scenario.}
\begin{IEEEproof}
\yet{To prove this lemma, we need to show that for any network with legitimate transmission pair density $\lambda_T$ and eavesdropper density $\lambda_E$, we can properly set the parameters $\lambda_P$ and $\rho$ of the proposed SCJ scheme to achieve a better STC performance than the case without applying the SCJ scheme.
Obviously, one  such setting is that $\rho=0$ and $\lambda_P$ can be any value.
In this case, no potential jammers inside the LoS balls of the receivers become jammers. 
In other words, all the jammers lie outside the LoS balls of the receivers.
Thus, we have $\lambda_J=\rho p_N \lambda_P=0$.
For a typical receiver, the jammers have NLoS links to it and are at least distance $D$ away, so their interference can be neglected due to the severe signal attenuation. 
This is equivalent to the case without the SCJ scheme where there is no interference from jammers.
Hence, the connection probability in this case is identical to that of the case without the SCJ scheme.
However, this is not the case for eavesdroppers, since the jammers may be inside their LoS balls and have LoS links to them.
Thus, the eavesdroppers in this case suffer from more interference than in the case without the SCJ scheme, leading to a greater secrecy probability.
As a result, the proposed SCJ scheme improves the network STC performance.
}
\end{IEEEproof}
\end{lemma}

The secrecy probabilities for the PJ case is also given in the following corollary based on Theorem \ref{theorem:ps-general-case}.
\begin{corollary}
The secrecy probability for the transmission pairs under the PJ scheme can be given by 
\begin{IEEEeqnarray}{rCl}
p_s^{\mathrm{PJ}}&\approx&\exp\Bigg(-2\pi \sum_{\mathsf b}\sum_{\hat{\mathsf g}}p_{\mathsf b}q_{\hat{\mathsf g}}\lambda_E \!\!\sum_{k=1}^{N_{\mathsf b}}\binom{N_{\mathsf b}}{k}(-1)^{k+1}\\
&&\ \ \ \int_{0}^De^{-k\nu_{\mathsf b} r_e^{\alpha_{\mathsf b}}\frac{\sigma^2}{ P}}\mathcal L_{I_{\Phi_J}^z}(\lambda_T+ \varrho\lambda_P, k \nu_{\mathsf b} r_e^{\alpha_{\mathsf b}})r_e\mathrm d r_e\Bigg)\nonumber\\
&&\times\exp\Bigg(-2\pi\sum_{\hat{\mathsf g}} q_{\hat{\mathsf g}}\lambda_E \sum_{k=1}^{N_{N}}\binom{N_{N}}{k}(-1)^{k+1}\nonumber\\
&&\int_{D}^\infty e^{-k\nu_{N} r_e^{\alpha_{N}}\frac{\sigma^2}{ P}}\mathcal L_{I_{\Phi_J}^z}(\lambda_T+ \varrho\lambda_P, k \nu_{N} r_e^{\alpha_{N}})r_e\mathrm d r_e\Bigg), \nonumber
\end{IEEEeqnarray}
where $\nu_{\mathsf b} = \frac{\tau_{\mathsf b}(2^{R_e}-1)}{\hat{\mathsf g}}$, $\tau_{\mathsf b}=N_{\mathsf b}(N_{\mathsf b}!)^{-1/N_{\mathsf b}}$ and $\mathcal L_{I_{\Phi_J}^z}(\cdot, \cdot)$ is given by Lemma \ref{lemma:LT-Phi-J-E-simple-case}.
\end{corollary}
\begin{IEEEproof}
The proof follows directly from Theorem \ref{theorem:ps-general-case}.
\end{IEEEproof}

\begin{remark}
\yet{Note that deriving the accurate results on secrecy probability is extremely challenging, because this involves the complicated interference modeling for legitimate receivers and eavesdroppers under the random point process (in particular the Poisson Hole Process concerned in this paper), which remains a long-lasting open problem by now \cite{haenggi2012stochastic}. That is why the available works (including this work) mainly focus on the efficient approximation for interferences at legitimate receivers and eavesdroppers and thus deriving bounds on the secrecy probability, see, for example, \cite{YZhu2017TWC,WangC2016TWC,SVuppala2018TCOM, WWang2018TCOM, XSun2019TIFS}.}
\end{remark}
%\begin{remark}
%\yet{Although the proofs of the lemmas and theorems follow the fundamental and standard frameworks in the literature, the derivations of the Laplace transforms of interference to legitimate receivers and eavesdroppers significantly differ from recent papers.}
%\end{remark}

\subsection{Optimal SCJ Parameters} \label{sec:opt_scj_pmts}
The most important parameter in the proposed SCJ scheme is the interference-control factor $\rho$, i.e., the jammer selection probability inside the region $\mathcal B$ covered by the LoS balls of the receivers. 
In general, the noise generated by the jammers is helpful to suppress eavesdroppers but harmful to the receivers. 
Thus, the larger the parameter $\rho$ is, the larger the secrecy probability $p_s$ is but the smaller the connection probability $p_c$ becomes. 
As a result, there exists an optimal $\rho$, denoted by $\rho^*$, that maximizes the STC $\bar{R}_s(\rho)$. 
Thus, we can formulate the following optimization problem to find the $\rho^*$, provided all the other system parameters (e.g., $\lambda_T$, $\lambda_E$ and $\lambda_P$) are given.
\begin{IEEEeqnarray}{rCl}
\mathbf{P1}:\	\rho^*&=&\underset{\rho\in[0,1]}{\argmax} \	p_c(\rho) p_s(\rho) (R_t-R_e)\lambda_T.
\end{IEEEeqnarray}
Similarly, there also exists an optimal $\varrho$, denoted by $\varrho^*$, to maximize the STC under the PJ scheme. 
The $\varrho^*$ can be obtained by solving the following optimization problem.
\begin{IEEEeqnarray}{rCl}
\mathbf{P2}:\	\varrho^*&=&\underset{\varrho\in[0,1]}{\argmax} \	p_c^{\mathrm{PJ}}(\varrho) p_s^{\mathrm{PJ}}(\varrho) (R_t-R_e)\lambda_T.
\end{IEEEeqnarray}
\yet{Network designers may be also interested in the optimal network STC performance under a constraint $\varepsilon$ on the total density of legitimate nodes (i.e., legitimate pairs and potential jammers). 
Thus, we also formulate the following two optimization problems to facilitate network design.
\begin{IEEEeqnarray}{rCl}
\mathbf{P3}:\underset{\rho\in[0,1], \lambda_T+\lambda_P\in[0,\varepsilon]}{\max} \	p_c(\rho) p_s(\rho) (R_t-R_e)\lambda_T,
\end{IEEEeqnarray}
\begin{IEEEeqnarray}{rCl}
\mathbf{P4}:\underset{\varrho\in[0,1], \lambda_T+\lambda_P\in[0,\varepsilon]}{\max} \	p_c^{\mathrm{PJ}}(\varrho) p_s^{\mathrm{PJ}}(\varrho) (R_t-R_e)\lambda_T.
\end{IEEEeqnarray}
Although closed-form solutions to these four optimization problems may not be available, they can be numerically solved.}

\section{Numerical Results} \label{sec:num-res}  
This section provides numerical results to evaluate the optimal STC performance achieved by the proposed SCJ scheme. 
We adopt the parameter settings in Table \ref{tb:pmts} unless stated otherwise, and also set  the codeword rates as $R_t=8$ bps/Hz, $R_e=4$ bps/Hz and the factor $\beta$ as $\beta=0.8$ for all the figures. 
All results are obtained by numerically solving the optimization problems in Section \ref{sec:opt_scj_pmts}.

\subsection{Optimal STC vs. Density of Potential Jammers $\lambda_P$}
\begin{figure}[h]
 \centering
 \includegraphics[width=0.45\textwidth]{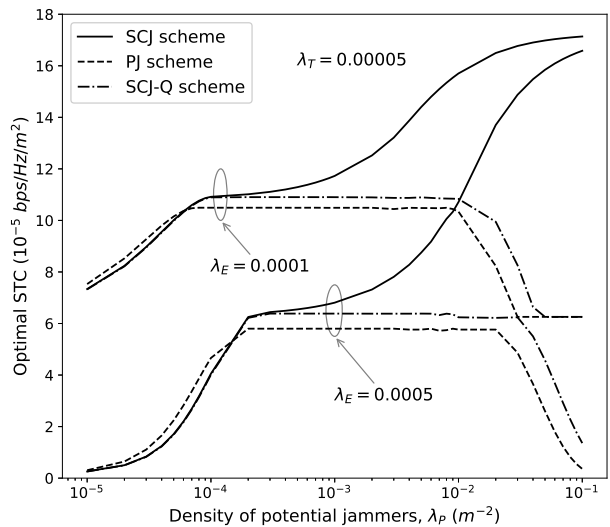}
 \caption{Optimal STC vs. density of potential jammers $\lambda_P$.}
 \label{fig:optrsVsLP}
\end{figure}

\begin{figure}[h]
 \centering
 \includegraphics[width=0.45\textwidth]{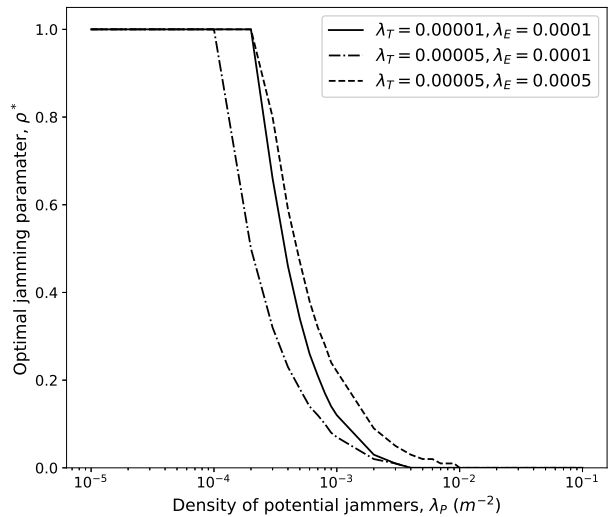}
 \caption{Optimal jamming parameter $\rho^*$ vs. density of potential jammers $\lambda_P$.}
 \label{fig:optrhoVsLP}
\end{figure}

Fig.~\ref{fig:optrsVsLP} shows the impacts of the density of potential jammers $\lambda_P$ on the optimal STC under both the SCJ and PJ  schemes. 
In this figure, we fix $\lambda_T$ as $0.00007$ and consider two cases with different density of eavesdroppers $\lambda_E$, i.e., ($\lambda_E =0.0001$ and $\lambda_E=0.0001$). 
We can see from Fig.~\ref{fig:optrsVsLP} that, for the SCJ scheme scheme, the optimal STC first increases as $\lambda_P$ increases and finally remains unchanged. 
This implies that the proposed SCJ scheme can can improve the network STC performance by deploying more jammers 
The reason for the finally unchanged optimal STC under the SCJ scheme is that, when $\lambda_P$ exceeds some threshold, the optimal jamming parameter $\rho^*$ decreases to $0$. 
This means that all the jammers are outside the LoS balls, i.e., the region $\mathcal B$, and thus generate interference only to the eavesdroppers. 
As a result, as $\lambda_P$ further increases from the threshold, the connection probability remains constant, while the secrecy probability will finally increase to one and stay unchanged, leading to a constant STC.
Different from the SCJ scheme,  as $\lambda_P$ keeps increasing, the optimal STC of the PJ scheme further decreases and finally remains constant. 
The differences between the behaviors of the two schemes show the benefits of the jammers outside the LoS balls of legitimate receivers (i.e., the jammers in $\Psi_J$) in improving the STC performance.
\yet{To demonstrate this expectation, we consider a variant SCJ scheme, named SCJ-Q scheme, where the jammers in $\Psi_J$ remain quiet, i.e., $\bar{\lambda}_J$ denotes the density of quiet jammers.  
We compare the SCJ-Q scheme and the SCJ scheme in terms of the optimal STC performance in Fig.  \ref{fig:optrsVsLP}. 
We can observe from Fig. \ref{fig:optrsVsLP} that the proposed SCJ scheme achieves better STC performance than the SCJ-Q scheme, implying that the jammers in $\Psi_J$ have a significant impact on the STC performance enhancement.}

A careful observation in Fig.~\ref{fig:optrsVsLP} indicates that when $\lambda_P$ is smaller than some value (about $0.00009$ in Fig.~\ref{fig:optrsVsLP}), the optimal STC achieved by the SCJ scheme is slightly smaller than that achieved by the PJ scheme. 
However,  when $\lambda_P$ is larger than this value, the SCJ scheme achieves better STC performance than the PJ scheme, and the gap between these two schemes increases significantly as $\lambda_P$ increases. 
This implies that the proposed SCJ scheme can greatly improve the STC performance achieved by the PJ scheme by deploying a large number of potential jammers, showing the effectiveness of the SCJ scheme in ensuring the secrecy of transmissions at the physical layer.  

Fig.~\ref{fig:optrhoVsLP} shows the optimal parameter $\rho^*$ of the SCJ scheme versus $\lambda_P$ under the three cases considered in Fig.~\ref{fig:optrsVsLP}. 
We can observe from Fig.~\ref{fig:optrhoVsLP} that $\rho^*$ decreases as  $\lambda_P$ increases, and finally decreases to $0$. 
This observation explains the reason for the finally unchanged STC in Fig.~\ref{fig:optrsVsLP}, and also serve as a guideline for setting the value of the jamming parameter $\rho$ to achieve the optimal STC. 
Another careful observation from Fig.  \ref{fig:optrhoVsLP} indicates that $\rho^*$ decreases as $\lambda_T$ increases while increases as $\lambda_E$ increases. 

\subsection{Optimal STC vs. Density of Transmission Pairs $\lambda_T$}
\begin{figure}[h]
 \centering
 \includegraphics[width=0.45\textwidth]{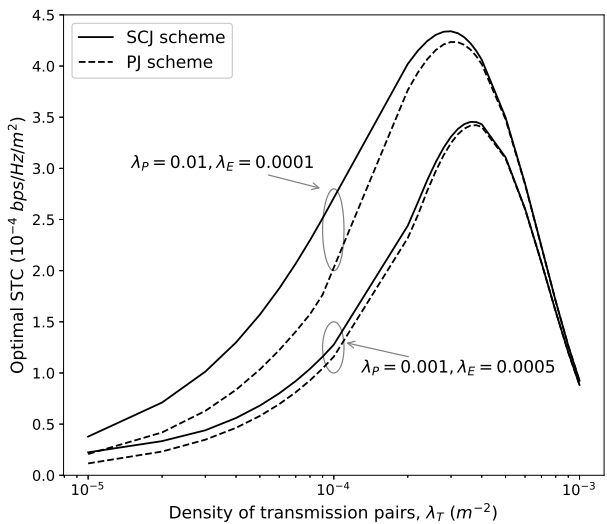}
 \caption{Optimal STC vs. density of transmission pairs $\lambda_T$.}
 \label{fig:optrsvsLT}
\end{figure}

\begin{figure}[h]
 \centering
 \includegraphics[width=0.45\textwidth]{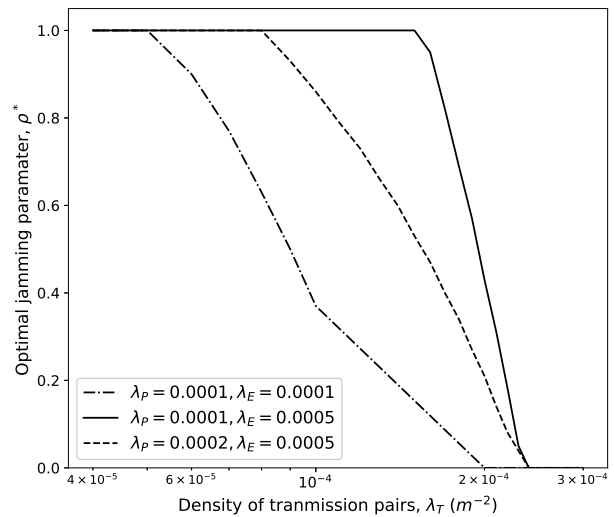}
 \caption{Optimal jamming parameter $\rho^*$ vs. density of transmission pairs $\lambda_T$.}
 \label{fig:optrhoVsLT}
\end{figure}

To investigate the impacts of $\lambda_T$  on the optimal STC performance, we show in Fig.~\ref{fig:optrsvsLT} the optimal STC versus $\lambda_T$ under two cases with different values of $\lambda_P$ and $\lambda_E$, i.e., ($\lambda_P=0.01$, $\lambda_E=0.0001$) and ($\lambda_P=0.001$, $\lambda_E=0.0005$). 
We can see from Fig.~\ref{fig:optrsvsLT} that as $\lambda_T$ increases, the optimal STC first increases, then drastically decreases and will finally vanish. 
This is because when $\lambda_T$ is small, the change of $\lambda_T$ dominates the trend of the optimal STC more than the changes of secrecy probability and connection probability do. 
However, when $\lambda_T$ is large, the network becomes dense and the interference from concurrent transmitters leads to significantly decreased connection probability. 
In this case, the decrease of the connection probability dominates the trend of the optimal STC more than the increase of the $\lambda_T$ does. 

Another observation from Fig.~\ref{fig:optrsvsLT} suggests that the STC gap between the SCJ scheme and the PJ scheme is large when $\lambda_T$ is small, and this gap vanishes as $\lambda_T$ increases. 
This is because smaller $\lambda_T$ yields smaller LoS ball region $\mathcal B$ and thus more jammers that are outside the $\mathcal B$. 
In addition, smaller $\lambda_T$ yields larger associated regions (as shown in Fig.~\ref{fig:region-approx}) inside the LoS balls of legitimate receivers. 
As a result, the jamming signals from the jammers are more detrimental to the eavesdroppers than to the legitimate receivers. 
Thus, the SCJ scheme outperforms the PJ scheme in this case. However, as $\lambda_T$ increases, the region $\mathcal B$ becomes larger and will finally cover the whole network region. 
In this case, the associated region inside the LoS ball of each legitimate receiver vanishes. 
Thus, the SCJ scheme is equivalent to the PJ scheme and thus achieves the same STC performance. 
The large gap between the SCJ and PJ schemes for small $\lambda_T$ indicates that the SCJ scheme is more effective when the density of transmission pairs is low. 

Fig.~\ref{fig:optrhoVsLT} shows the behavior of the optimal jamming parameter $\rho^*$ versus $\lambda_T$ under three cases, i.e., ($\lambda_P=0.0001$, $\lambda_E=0.0001$), ($\lambda_P=0.0001$, $\lambda_E=0.0005$) and ($\lambda_P=0.0002$, $\lambda_E=0.0005$).  
It can be seen from  Fig.~\ref{fig:optrhoVsLT} that the  $\rho^*$ decreases as $\lambda_T$ increases, which is consistent with the observation from Fig.~\ref{fig:optrhoVsLP}. 
This can also help us determine the optimal settings of the jamming parameter $\rho$ during the system design.

\subsection{\yet{Optimal STC and Energy Efficiency vs. Total Density Constraint $\varepsilon$}}
\begin{figure}[h]
 \centering
 \includegraphics[width=0.45\textwidth]{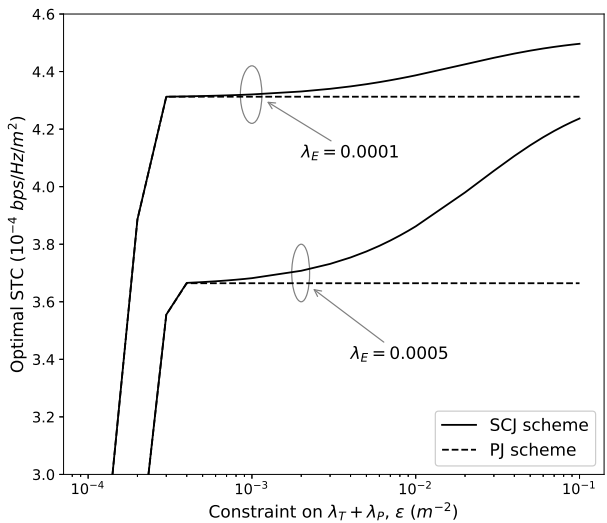}
 \caption{Optimal STC vs. total density constraint $\varepsilon$.}
 \label{fig:optrsvsepsilon}
\end{figure}
\yet{This subsection explores the impacts of the total density constraint $\varepsilon$ on the optimal STC performance. 
By solving optimization problems P3 and P4, we show in Fig. \ref{fig:optrsvsepsilon} the optimal STC versus $\varepsilon$ under two cases of $\lambda_E=0.0001$ and $\lambda_E=0.0005$.
The results show that the optimal STC increases as $\varepsilon$ increases for both the proposed SCJ scheme and the PJ scheme.
A careful observation indicates that  the optimal STC of the PJ scheme finally remains unchanged, while that of the proposed SCJ scheme keeps increasing, 
yielding an increasingly large STC gap.
This suggests that the proposed SCJ scheme outperforms the PJ scheme in terms of the STC performance under the total density constraint, and the performance gap between these two schemes enlarges as the constraint  $\varepsilon$ increases.}

\begin{figure}[h]
 \centering
 \includegraphics[width=0.45\textwidth]{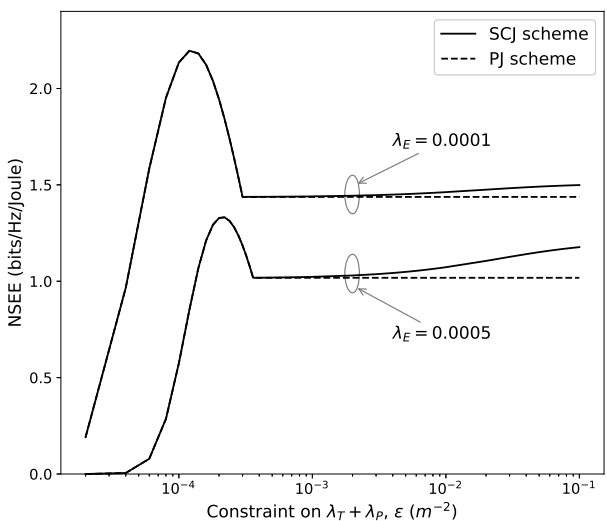}
 \caption{NSEE vs. total density constraint $\varepsilon$.}
 \label{fig:nseevsepsilon}
\end{figure}

\yet{The better STC performance of the SCJ is achieved by deploying more potential jammers, which may consume more energy.
Thus, we need to compare the energy efficiency of both schemes. 
To do this, we adopt the metric of network-wide energy efficiency (NSEE) from \cite{TZheng2017TWC}, which can be formulated as
\begin{IEEEeqnarray}{rCl}
NSEE=\frac{\bar{R}_s}{(\lambda_T + \lambda_J+e^{-\lambda_R \pi D^2}\bar{\lambda}_J)P},
\end{IEEEeqnarray}
where $\bar{R}_s$ denotes the STC and the denominator denotes the average power per unit area, including that of the legitimate transmitters, jammers in $\Phi_J$ and jammers in $\Psi_J$.
The unit of the NSEE is $bits/Hz/Joule$.
For NSEE comparison, we consider the case in Fig. \ref{fig:optrsvsepsilon} where both schemes achieve the optimal STC under the total constraint $\varepsilon$.
Fig. \ref{fig:nseevsepsilon} shows the NSEE of both schemes versus $\varepsilon$ under two  cases of $\lambda_E=0.0001$ and $\lambda_E=0.0005$.
The results show that the NSEEs of both schemes first increase and then decrease as $\varepsilon$ increases.
Similar to the STC, the NSEE of the PJ scheme finally remains unchanged, while that of the SCJ scheme continues increasing.
This indicates that the SCJ scheme also outperforms the PJ scheme in terms of the NSEE performance and the performance gap between these two schemes enlarges as the constraint   $\varepsilon$ increases. 
The results in this subsection show the superiority of the proposed SCJ scheme over the PJ scheme in terms of not only the STC performance but also energy efficiency.}

\subsection{Optimal STC vs. Density of Eavesdroppers $\lambda_E$}
\begin{figure}[h]
 \centering
 \includegraphics[width=0.45\textwidth]{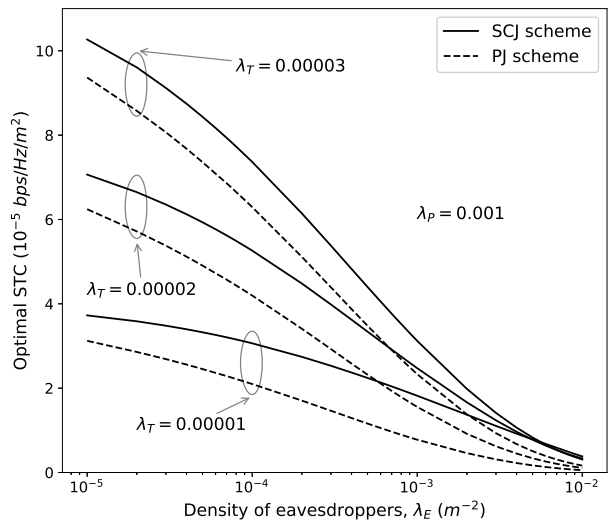}
 \caption{Optimal STC vs. density of eavesdroppers $\lambda_E$.}
 \label{fig:optrsvsLE}
\end{figure}

\begin{figure}[h]
 \centering
 \includegraphics[width=0.45\textwidth]{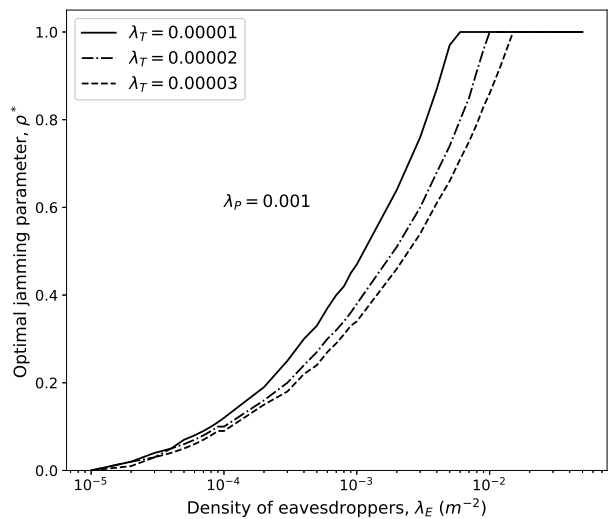}
 \caption{Optimal jamming parameter $\rho^*$ vs. density of eavesdroppers $\lambda_E$.}
 \label{fig:optrhovsLE}
\end{figure}

To explore the impacts of density of eavesdroppers $\lambda_E$ on the STC performance, we show the optimal STC versus $\lambda_E$ in Fig.~\ref{fig:optrsvsLE} under the setting of $\lambda_P=0.001$. Three cases of $\lambda_T$ are considered, which are $\lambda_T=0.00001$, $\lambda_T=0.00002$ and $\lambda_T=0.00003$. We can observe from Fig.~\ref{fig:optrsvsLE} that the optimal STC decreases as $\lambda_E$ increases. This is intuitive since more eavesdroppers leads to lower secrecy probability and thus smaller STC. The superiority of the SCJ scheme over the PJ scheme in terms of the STC performance can also be seen from the figure. Fig.~\ref{fig:optrhovsLE} illustrates the behavior of the optimal jamming parameter $\rho^*$ versus $\lambda_E$ under the same settings of $\lambda_T$ and $\lambda_P$ as adopted in Fig.~\ref{fig:optrsvsLE}. As seen from the figure, the $\rho^*$ increases as $\lambda_E$ increases. This implies that when there exists more eavesdroppers in the network, we need to adopt a larger jamming parameter $\rho$ to achieve the optimal STC performance.

\section{Conclusions} \label{sec:con} 
A Sight-based Cooperative Jamming (SCJ) scheme was proposed in this paper to improve the physical layer security performance in millimeter-wave (mmWave) ad hoc networks, and the related theoretical framework was also developed to model the secrecy transmission capacity (STC) achievable from adopting the jamming scheme. 
The results in this paper indicate that the proposed jamming scheme can significantly improve the STC of mmWave ad hoc networks, especially for networks with low density of transmission pairs. 
For an SCJ-based ad hoc network with given densities of transmission pairs, potential jammers and eavesdroppers, our theoretical framework can serve as a guideline on the proper settings of jamming parameter such that the optimal STC performance of the network can be achieved. 
\yet{Although this paper considers the “nearest receiver” association policy, other policies are also possible, like the “smallest average received power” policy. Thus, an interesting topic is the design of optimal association policy, which serves as one future work.}

\appendices
\section{Proof of Lemma \ref{lemma:LT-Phi-J-simple-case}}\label{app:LT-Phi-J-simple-case}  
Due to the independent thinning by the antenna gains between the jammers and $y_0$, the jammers in $\Phi_J$ seen from $y_0$ can be modeled by independent sub-PPPs $\Phi_J^{\mathsf g}$ with density $q_{\mathsf g} \lambda_{J}$, where $\mathsf g\in\{G_TG_R, G_Tg_R, g_TG_R, g_Tg_R\}$ denotes the antenna gain. Thus, we have 
$I_{\Phi_J}^{y_0}=\sum_{\mathsf g}I_{\Phi_{J}^{\mathsf g}}^{y_0}$, where 
\begin{align}
I_{\Phi_J^{\mathsf g}}^{y_0}=\sum_{x\in\Phi_{J}^{\mathsf g}} \mathsf g  h^N_{x,y_0} \lVert x-y_0 \rVert^{-\alpha_N}.
\end{align}

Since the sub-PPPs $\Phi_{J}^{\mathsf g}$ are independent, we have  
\begin{IEEEeqnarray}{rCl}\label{eqn:LT-Phi-J-simple-case-proof-1} 
\mathcal L_{I_{\Phi_J}^{y_0}}(s)&=&\prod_{\mathsf g}\mathcal L_{I_{\Phi_J^{\mathsf g}}^{y_0}}(s). 
\end{IEEEeqnarray} 
Placing $y_0$ at the origin (i.e., $y_0 = o$F) yields 
\begin{IEEEeqnarray}{rCl}\label{eqn:LT-Phi-J-simple-case-proof-2}
\mathcal L_{I_{\Phi_{J}^{\mathsf g}}^{y_0}}(s)&=&\mathbb E\Big[\exp\Big(-s\sum_{x\in \Phi_{J}^{\mathsf g}}\mathsf g h^N_{x,o}\lVert x\rVert^{-\alpha_N}\Big)\Big]\nonumber\\ 
&\overset{(a)}=&\mathbb E_{\Phi_{J}^{\mathsf g}}\Big[\prod_{x\in \Phi_{J}^{\mathsf g}}\mathbb E_{h^N_{x,o}}\Big[e^{-s\mathsf g h^N_{x,o} \lVert x\rVert^{-\alpha_N}}\Big]\Big]\nonumber\\
&\overset{(b)}=&\mathbb E_{\Phi_{J}^{\mathsf g}}\Big[\prod_{x\in \Phi_{J}^{\mathsf g}}\Big(1+\frac{s\mathsf g}{N_N \lVert x\rVert^{\alpha_N}}\Big)^{-N_N}\Big]\nonumber\\
&\overset{(c)}=&\exp{\Big(-q_{\mathsf g} \lambda_J\int_{\mathbb R^2}F_N(s\mathsf g, \lVert x\rVert)\mathrm dx\Big)}\nonumber\\
&\overset{(d)}=&\exp{\Big(-2\pi q_{\mathsf g} \lambda_J\int_0^\infty F_N(s\mathsf g, r)r\mathrm dr\Big)},
\end{IEEEeqnarray}
where $F_{\mathsf b}(\cdot, \cdot)$ is given by (\ref{eqn:Fb}), $(a)$ follows since $h^N_{x,o}$ are i.i.d.,  $(b)$ follows from the moment generating function (MGF) of the gamma random variable $h^N_{x,o}$, $(c)$ follows after applying the probability generating functional (PGF) of the PPP \cite{Chiu2013} and $(d)$ follows after changing to the polar coordinates.  Substituting (\ref{eqn:LT-Phi-J-simple-case-proof-2}) into (\ref{eqn:LT-Phi-J-simple-case-proof-1}) completes the proof.

\section{Proof of Lemma \ref{lemma:LT-Psi-J-E-simple-case}}\label{app:LT-Psi-J-E-simple-case}  
We divide $\Psi_J$ into sub-point processes inside and outside the LoS ball $B(z,D)$ based on the independent thinning by link condition and antenna gains. 
Thus, we have
\begin{IEEEeqnarray}{rCl}\label{eqn:LT-Psi-J-E-simple-case-1}
\mathcal L_{I_{\Psi_J}^z}(s, r_e)&=&\prod_{\mathsf b}\prod_{\hat{\mathsf g}}\mathcal L_{I_{\Psi_{J,I}^{\mathsf b,\hat{\mathsf g}}}^z}(s, r_e)\prod_{\hat{\mathsf g}}\mathcal L_{I_{\Psi_{J,O}^{N,\hat{\mathsf g}}}^z}(s, r_e),
\end{IEEEeqnarray}
where $\Psi_{J,I}^{\mathsf b,\hat{\mathsf g}}$ (resp. $\Psi_{J,O}^{N,\hat{\mathsf g}}$) denotes the sub-PPPs inside (resp. outside) $B(z,D)$.
Note that $\Psi_{J,I}^{\mathsf b,\hat{\mathsf g}}$ is equivalent to $\bar{\Phi}_{J,I}^{\mathsf b,\hat{\mathsf g}}\cap B^c(y_0,D)$, where $\bar{\Phi}_{J,I}^{\mathsf b,\hat{\mathsf g}}$ is the homogeneous baseline PPP with density $p_{\mathsf b}q_{\hat{\mathsf g}}\bar{\lambda}_J$ inside $B(z,D)$. Assuming that the eavesdropper $z$ is located at the origin and defining $\Xi_I=B^c(y_0,D)\cap B(z,D)$ and $\Xi_O=B^c(y_0,D)\cap B^c(z,D)$, we have
\begin{IEEEeqnarray}{rCl}\label{eqn:LT-Psi-J-E-simple-case-2}
\mathcal L_{I_{\Psi_{J,I}^{\mathsf b,\hat{\mathsf g}}}^z}(s,r_e)&=&\mathbb E\Big[\exp\Big(-s\sum_{x\in \bar{\Phi}_{J}^{\mathsf b,\hat{\mathsf g}}}\hat{\mathsf g} h^{\mathsf b}_{x,o}\lVert x\rVert^{-\alpha_{\mathsf b}}\Big)\Big]\nonumber\\
&=&\exp\Big(\!\!-\!\!p_{\mathsf b}q_{\hat{\mathsf g}}\bar{\lambda}_J\underbrace{\int_{\Xi_I}F_{\mathsf b}(s\hat{\mathsf g}, \lVert x\rVert)\mathrm dx}_{T_1}\!\!\Big),
\end{IEEEeqnarray} 
\begin{IEEEeqnarray}{rCl}\label{eqn:LT-Psi-J-E-simple-case-3}
\mathcal L_{I_{\Psi_{J,O}^{N,\hat{\mathsf g}}}^z}(s,r_e)&=&\mathbb E\Big[\exp\Big(-s\sum_{x\in \bar{\Phi}_{J,O}^{N,\hat{\mathsf g}}}\hat{\mathsf g} h^{N}_{x,o}\lVert x\rVert^{-\alpha_{N}}\Big)\Big]\nonumber\\
&=&\exp\Big(\!\!-\!\!q_{\hat{\mathsf g}}\bar{\lambda}_J\underbrace{\int_{\Xi_O}F_{N}(s\hat{\mathsf g}, \lVert x\rVert)\mathrm dx}_{T_2}\!\!\Big),
\end{IEEEeqnarray} 
We can see that $T_1$ and $T_2$ depend on the distance $\lVert y_0\rVert$ between $y_0$ and $z=o$. 
If $\lVert y_0\rVert \in [0,D)$, $T_1$ and $T_2$ can be given by 
\begin{IEEEeqnarray}{rCl}\label{eqn:T-1-1}
T_1\!&=&\!2\int_{D\!-\!\lVert y_0\rVert}^{D}\!\!\!\!F_{\mathsf b}(s\hat{\mathsf g},r)\!\Big(\!\pi \!\!-\!\! \arccos(\!\frac{\lVert y_0\rVert^2\!\!+\!\!r^2\!\!-\!\!D^2}{2\lVert y_0\rVert r})\!\Big)r\mathrm d r,
\end{IEEEeqnarray}
\begin{IEEEeqnarray}{rCl}\label{eqn:T-2-1}
T_2\!&=&2\pi \int_{D}^{\infty}\!\!\!\!F_{N}(s\hat{\mathsf g},r)r\mathrm d r\\
&&-\!2\int_{D}^{\lVert y_0\rVert+D}\!\!\!\!F_{N}(s\hat{\mathsf g},r) \arccos(\!\frac{\lVert y_0\rVert^2\!\!+\!\!r^2\!\!-\!\!D^2}{2\lVert y_0\rVert r})r\mathrm d r.\nonumber
\end{IEEEeqnarray}
If $\lVert y_0\rVert \in [D,2D)$, $T_1$ can be given by
\begin{IEEEeqnarray}{rCl}\label{eqn:T-1-2}
T_1&=&2\pi\int_{0}^{D}F_{\mathsf b}(s\hat{\mathsf g},r)r\mathrm d r\\
&&-2\int_{\lVert y_0\rVert-D}^{D}\!\!\!\! F_{\mathsf b}(s\hat{\mathsf g},r)\arccos(\frac{\lVert y_0\rVert^2\!+\!r^2\!-\!D^2}{2\lVert y_0\rVert r})r\mathrm dr,\nonumber
\end{IEEEeqnarray}
$T_2$ is identical to that of the case of $\lVert y_0\rVert \in [0,D)$.
If $\lVert y_0\rVert\in [2D, \infty)$, $T_1$ and $T_2$ can be given by 
\begin{IEEEeqnarray}{rCl}\label{eqn:T-1-3}
T_1&=&2\pi\int_{0}^{D}F_{\mathsf b}(s\hat{\mathsf g},r)r\mathrm d r,
\end{IEEEeqnarray}
\begin{IEEEeqnarray}{rCl}\label{eqn:T-2-3}
T_2&=&2\pi\int_{D}^{\infty}F_{N}(s\hat{\mathsf g},r)r\mathrm d r\\
&&-2\int_{\lVert y_0\rVert-D}^{\lVert y_0\rVert+D}F_{N}(s\hat{\mathsf g},r)\arccos(\!\frac{\lVert y_0\rVert^2\!\!+\!\!r^2\!\!-\!\!D^2}{2\lVert y_0\rVert r})r\mathrm d r\nonumber
\end{IEEEeqnarray}
Summarizing (\ref{eqn:T-1-1}), (\ref{eqn:T-1-2}) and (\ref{eqn:T-1-3}) yields the $T_1(s,\lVert y_0\rVert)$ in (\ref{eqn:T_1}), and summarizing (\ref{eqn:T-2-1}) and (\ref{eqn:T-2-3}) yields the $T_2(s,\lVert y_0\rVert)$ in (\ref{eqn:T_2}). 
Note that $\lVert y_0\rVert\in[\lvert r_e-r_0\rvert, r_e+r_0]$ and the probability density function (PDF) of $\lVert y_0\rVert$ can be given by (\ref{eqn:pdf-h}). 
Thus, substituting (\ref{eqn:LT-Psi-J-E-simple-case-2}) and (\ref{eqn:LT-Psi-J-E-simple-case-3}) into (\ref{eqn:LT-Psi-J-E-simple-case-1}) and then taking the expectation of (\ref{eqn:LT-Psi-J-E-simple-case-1}) in terms of $\lVert y_0\rVert$ completes the proof.

\section{Proof of Theorem \ref{theorem:ps-simple-case}}\label{app:theorem-ps-simple-case}
Assuming $x_0$ is located at the origin, we have
\begin{IEEEeqnarray}{rCl}\label{eqn:ps-simple-case-1} 
&&\mathbb P\Big(\cap_{z\in \Phi_{E,I}^{\mathsf b,\hat{\mathsf g}}} \mathrm{SINR}_{x_0,z}^{\mathsf b,\hat{\mathsf g}}\leq 2^{R_e}-1\Big)\\
&\overset{(a)}=&\mathbb E\Big[\prod_{z\in \Phi_{E,I}^{\mathsf b,\hat{\mathsf g}}}\mathbb P\Big(h_{o,z}^{\mathsf b}\!\leq\! (2^{R_e}\!\!-\!\!1)\lVert z\rVert^{\alpha_{\mathsf b}}(\sum_{\varphi}I_{\varphi}^z\!+\!\frac{\sigma^2}{P})/\hat{\mathsf g}\Big)\Big]\nonumber\\
&\overset{(b)}\geq&\mathbb E_{\{\varphi\}, \Phi_{E,I}^{\mathsf b,\hat{\mathsf g}}}\Big[\prod_{z\in \Phi_{E,I}^{\mathsf b,\hat{\mathsf g}}}\left(1-e^{- \nu_{\mathsf b} \lVert z\rVert^{\alpha_{\mathsf b}}(\sum_{\varphi}I_{\varphi}^z+\frac{\sigma^2}{P})}\right)^{N_{\mathsf b}}\Big]\nonumber\\ 
&\overset{(c)}=&\mathbb E_{\{\varphi\}}\Big[\exp\Big(-2\pi p_{\mathsf b}q_{\hat{\mathsf g}}\lambda_E \nonumber\\
&&\int_{0}^D \Big[1- \Big(1-e^{-\nu_{\mathsf b} r_e^{\alpha_{\mathsf b}}(\sum_{\varphi}I_{\varphi}^z+\frac{\sigma^2}{P})}\Big)^{N_{\mathsf b}}\Big]r_e \mathrm d r_e\Big)\Big]\nonumber\\
&\overset{(d)}\geq&\exp\Big(-2\pi p_{\mathsf b}q_{\hat{\mathsf g}}\lambda_E \sum_{k=1}^{N_{\mathsf b}}\binom{N_{\mathsf b}}{k}(-1)^{k+1}\nonumber\\
&&\ \ \ \ \ \ \ \ \ \ \ \int_{0}^D e^{-k \nu_{\mathsf b} r_e^{\alpha_{\mathsf b}}\frac{\sigma^2}{ P}}\prod_{\varphi}\mathbb E_{\varphi}[e^{-k \nu_{\mathsf b} r_e^{\alpha_{\mathsf b}}I_{\varphi}^z}]r_e\mathrm d r_e\Big)\nonumber\\
&=&\exp\Big(-2\pi p_{\mathsf b}q_{\hat{\mathsf g}}\lambda_E \sum_{k=1}^{N_{\mathsf b}}\binom{N_{\mathsf b}}{k}(-1)^{k+1}\int_{0}^D\mathrm d r_e\nonumber\\
&&\ \ \  e^{-k\nu_{\mathsf b} r_e^{\alpha_{\mathsf b}}\frac{\sigma^2}{ P}}\mathcal L_{I_{\Phi_J}^z}( k \nu_{\mathsf b} r_e^{\alpha_{\mathsf b}})\mathcal L_{I_{\Psi_J}^z}( k \nu_{\mathsf b} r_e^{\alpha_{\mathsf b}}, r_e)r_e\Big)\nonumber,
\end{IEEEeqnarray}
where $(a)$ follows since $h^{\mathsf b}_{o,z}$ are i.i.d., $(b)$ follows from the Lemma $1$ in \cite{AThornburg2016TSP}, $(c)$ follows after applying the PGF of the PPP and $(d)$ follows from the binomial theorem and the Jensen's inequality. Similarly, we can prove that
\begin{IEEEeqnarray}{rCl}\label{eqn:ps-simple-case-2} 
&&\mathbb P\Big(\cap_{z\in \Phi_{E,O}^{N,\hat{\mathsf g}}} \mathrm{SINR}_{x_0,z}^{N,\hat{\mathsf g}}\leq 2^{R_e}-1\Big)\\
&&\geq \exp\Big(-2\pi q_{\hat{\mathsf g}}\lambda_E \sum_{k=1}^{N_N}\binom{N_N}{k}(-1)^{k+1}\int_{D}^\infty\mathrm d r_e\nonumber\\
&&\ \ \  e^{-k\nu_N r_e^{\alpha_N}\frac{\sigma^2}{ P}}\mathcal L_{I_{\Phi_J}^z}( k \nu_N r_e^{\alpha_N})\mathcal L_{I_{\Psi_J}^z}( k \nu_N r_e^{\alpha_N}, r_e)r_e\Big).\nonumber
\end{IEEEeqnarray}
Substituting (\ref{eqn:ps-simple-case-1}) and  (\ref{eqn:ps-simple-case-2}) into (\ref{eqn:ps-formulation-simple-case}) completes the proof. 

\section{Proof of Lemma \ref{lemma:LT-Phi-J-general-case}}\label{app:LT-Phi-J-general-case}
We first calculate $\xi$ as the ratio of the expected area of the non-associated region inside $A_2$ to the area of $A_2$. Note that the non-associated region in $A_2$ is the union of many sub-non-associated regions, each formed by a D2D receiver located in the region $\Xi=B(y_0,2D)\backslash B(y_0,V_2)$. To simplify the calculation, we ignore the overlaps of the sub-non-associated regions and calculate the area of the non-associated region as the sum of the areas of the sub-non-associated regions. Assuming $y_0=o$, the area of the sub-non-associated region formed by a D2D receiver $y\in\Xi$  is $A(\lVert y\rVert)$, where $A(\cdot)$ is given by (\ref{eqn:A}). Thus, the area of the non-associated region in $A_2$ is
$\sum_{y\in\Phi_R\cap \Xi }A(\lVert y\rVert)$.
Applying the Campbell theorem \cite{Chiu2013}, we obtain the expected area
\begin{align}
\mathbb E_{\Phi_R}\Big[\sum_{y\in\Phi_R\cap \Xi}A(\lVert y\rVert)\Big]=2\pi \lambda_R\int_{V_2}^{2D}A(r)r\mathrm dr.
\end{align}
Since the area of the non-associated region must be no larger than that of $A_2$, which is $\pi (D^2-(V_2/2)^2)$, $\xi$ can be given by the $\xi(V_2)$ in (\ref{eqn:varrho}).  

Next, we calculate the Laplace transform of the interference from the three regions $A_1$, $A_2$ and $A_3$. Based on Lemma \ref{lemma:LT-Phi-J-simple-case}, we can derive the three Laplace transforms as 
\begin{IEEEeqnarray}{rCl}\label{eqn:LT-A1}
\mathcal L_{A_1}&=&\exp\!\Big(\!-\lambda_J\sum_{\mathsf b}\sum_{\mathsf g}p_{\mathsf b}q_{\mathsf g} \int_{A_1} F_{\mathsf b}(s\mathsf g, \lVert x\rVert)\mathrm dx\Big),
\end{IEEEeqnarray}
\begin{IEEEeqnarray}{rCl}\label{eqn:LT-A2}
\mathcal L_{A_2}&\approx&\exp\!\Big(\!-\xi(V_2) p_{L}\lambda_J\sum_{\mathsf g}q_{\mathsf g} \int_{A_2} F_L(s\mathsf g, \lVert x\rVert)\mathrm dx\Big)\\
&&\times\exp\Big(-(1\!-\!\xi(V_2)p_{L})\lambda_J\sum_{\mathsf g}q_{\mathsf g}\!\! \int_{A_2}\! F_N(s\mathsf g, \lVert x\rVert)\mathrm dx\Big)\nonumber,
\end{IEEEeqnarray}
and
\begin{IEEEeqnarray}{rCl}\label{eqn:LT-A3}
\mathcal L_{A_3}\!&=&\!\exp\!\Big(\!\!-\lambda_J\!\! \sum_{\mathsf g}\!q_{\mathsf g}\!\!\int_{B(y_0,D)\backslash(A_1\cup A_2)}\!\!\! \!F_N(s\mathsf g, \lVert x\rVert)\mathrm d x\!\Big).
\end{IEEEeqnarray}
Also, we need to calculate the Laplace transform of the interference from outside $B(y_0,D)$, which is
\begin{IEEEeqnarray}{rCl}\label{eqn:LT-A4}
\mathcal L_{A_4}\!&=&\!\exp\!\Big(\!\!-\lambda_J\!\! \sum_{\mathsf g}\!q_{\mathsf g}\!\!\int_{B^c(y_0,D)}\!\!\! \!F_N(s\mathsf g, \lVert x\rVert)\mathrm d x\!\Big).
\end{IEEEeqnarray}
Calculating the integrals in (\ref{eqn:LT-A1}), (\ref{eqn:LT-A2}) and (\ref{eqn:LT-A3}) in polar coordinates and then multiplying them yields 
\begin{IEEEeqnarray}{rCl}\label{eqn:LT-Phi-J-general-sub1}
\mathcal L_{I_{\Phi_J}^{y_0}}&=&\mathcal L_{A_1}\cdot \mathcal L_{A_2}\cdot \mathcal L_{A_3}\cdot \mathcal L_{A_4}\\
&\approx &\exp\Big(-2\pi\lambda_J\sum_{q_{\mathsf g}}q_{\mathsf g} \int_{0}^\infty F_N(s\mathsf g, r)r\mathrm dr \Big)\nonumber\\
&&\times\exp\Big(-p_{L}\lambda_J\sum_{q_{\mathsf g}}q_{\mathsf g} Q_1(s,V_1)\Big)\nonumber\\
&&\times\exp\Big(-\xi(V_2) p_{L}\lambda_J\sum_{q_{\mathsf g}}q_{\mathsf g}Q_2(s,V_1,V_2)\Big),\nonumber
\end{IEEEeqnarray}
where $Q1$ and $Q2$ are given by $(\ref{eqn:Q1})$ and $(\ref{eqn:Q2})$, respectively. Note that the joint PDFs of $V_1$ and $V_2$ can be given by (\ref{eqn:pdf-V1V2}) according to \cite{MOLTCHANOV20121146}. Finally, taking the expectation of $\mathcal L_{I_{\Psi_J}^{y_0}}$ in (\ref{eqn:LT-Phi-J-general-sub1}) in terms of $V_1$ and $V_2$ completes the proof.

\bibliographystyle{IEEEtran}
\bibliography{manuscript.bib}

\ifCLASSOPTIONcaptionsoff
  \newpage
\fi

\end{document}